\RequirePackage{fix-cm}
\RequirePackage{amsmath}
\documentclass[smallextended]{svjour3}
\smartqed 
\usepackage{balance}
\usepackage{colortbl}
\usepackage{graphicx}
\usepackage{lscape}
\usepackage{multirow}
\usepackage{tcolorbox}
\usepackage{url}
\usepackage{xspace}

\usepackage{tikz}
\usetikzlibrary{automata,arrows,positioning,calc}
\everymath{\displaystyle}

\usepackage{pgfplots}
\pgfplotsset{compat=1.15}

\renewcommand{\paragraph}[1]{\noindent\textbf{\textsf{#1}:}}

\newcommand{\ie}{\emph{i.e.,}\xspace}
\newcommand{\eg}{\emph{e.g.,}\xspace}

\newcommand{\etal}{\emph{et al.}\xspace}

\newcommand{\RQOne}{Do design patterns and--or design anti-patterns mutate during software evolution and what is the probability of the mutations?}

\newcommand{\RQTwo}{What types of changes lead to  mutations between design patterns and design anti-patterns?}

\newcommand{\RQThree}{What is the fault-proneness of mutated design patterns and anti-patterns and what transitions lead to more fault-prone mutations?}

\newcommand{\RQFour}{Do specific types of changes lead to increased fault-proneness during mutations?}

\begin{document}

\title{Investigating Design Anti-pattern and Design Pattern Mutations and Their Change- and Fault-proneness}

\titlerunning{Design Anti-pattern and Design Pattern Mutations}

\author{Zeinab (Azadeh) Kermansaravi \and
        Md Saidur Rahman \and
        Foutse Khomh 		\and
        Fehmi Jaafar \and
        Yann-Ga\"{e}l Gu\'{e}h\'{e}neuc
}
\institute{Zeinab Kermansaravi \at
            SWAT Lab, Ptidej Team, DGIGL, Polytechnique Montr\'{e}al, Montr\'{e}al, QC, Canada\\
		    \email{zeinab.kermansaravi@polymtl.ca}
	    \and
 		    Md Saidur Rahman \at
            SWAT Lab, DGIGL, Polytechnique Montr\'{e}al, Montr\'{e}al, QC, Canada\\
            \email{saidur.rahman@polymtl.ca}
    	\and
 		    Foutse Khomh \at
            SWAT Lab, DGIGL, Polytechnique Montr\'{e}al, Montr\'{e}al, QC, Canada\\
           	\email{foutse.khomh@polymtl.ca}
        \and
	    	Fehmi Jaafar \at
            Computer Research Institute of Montr\'{e}al, Montr\'{e}al, QC, Canada\\
            \email{fehmi.jaafar@crim.ca}
        \and	
            Yann-Ga\"{e}l Gu\'{e}h\'{e}neuc \at
            Ptidej Team, CSSE, Concordia University, Montr\'{e}al, QC, Canada\\
            \email{yann-gael.gueheneuc@concordia.ca}
}
\date{Received: date / Accepted: date}
\maketitle

\begin{abstract}
During software evolution, inexperienced developers may introduce design anti-patterns when they modify their software systems to fix bugs or to add new functionalities based on changes in requirements. Developers may also use design patterns to promote software quality or as a possible cure for some design anti-patterns. Thus, design patterns and design anti-patterns are introduced, removed, and mutated from one another by developers.

Many studies investigated the evolution of design patterns and design anti-patterns and their impact on software development. However, they investigated design patterns or design anti-patterns in isolation and did not consider their mutations and the impact of these mutations on software quality. Therefore, we report our study of bidirectional mutations between design patterns and design anti-patterns and the impacts of these mutations on software change- and fault-proneness. 

We analyzed snapshots of seven Java software systems with diverse sizes, evolution histories, and application domains. We built Markov models to capture the probability of occurrences of the different design patterns and design anti-patterns mutations. Results from our study show that (1) design patterns and design anti-patterns mutate into other design patterns and--or design anti-patterns. They also show that (2) some change types primarily trigger mutations of design patterns and design anti-patterns (renaming and changes to comments, declarations, and operators), and (3) some mutations of design anti-patterns and design patterns are  more faulty in specific contexts.

These results provide important insights into the evolution of design patterns and design anti-patterns and its impact on the change- and fault-proneness of software systems.

\end{abstract}

\section{Introduction}
\label{sec:introduction}

\subsection{Background and Motivation}

Quality assurance is one of the most critical challenges in software development and evolution \cite{gamma1995design}. Under tight deadlines and other business constraints, developers may take poor design or coding decisions and may follow bad practices. These poor design decisions and bad coding practices are collectively called ``smells'' \cite{stamelos2002code,khomh2007perception}.

Smells are divided into several different categories \cite{fowler1999refactoring}, such as code smells, design smells, lexical smells, etc. Code smells include low-level problems in the source code that may be symptoms of poor coding practices \cite{van2002java,mantyla2003taxonomy}. Design smells include design anti-patterns that describe poor solutions to recurring design problems. Design anti-patterns have been reported to make software development and evolution difficult \cite{khomh2012exploratory}. They affect program comprehension \cite{abbes2011empirical} and increase change- and fault-proneness \cite{khomh2012exploratory}. 

Opposite to design anti-patterns, design patterns are good solutions to recurring design problems, which promote code reuse and increase reliability, readability, and flexibility \cite{gamma1995design}. Gamma \etal \cite{gamma1995design} suggested using specific design patterns to ease evolution and increase reuse and flexibility. Studies showed that design patterns often increase the quality of software systems (\eg{} \cite{ampatzoglou2012methodology}). Yet, a few studies showed that some design patterns can decrease some quality attributes \cite{khomh2008design}. 

Recent studies \cite{jaafar2013analysing,vokavc2004defect,aversano2007empirical} showed the existence of relationships between design patterns and design anti-patterns. Such relationships can help developers to understand their systems better and simplify development and evolution \cite{jaafar2013analysing}. Yet, no study considered that design patterns sometimes (d)evolve into design anti-patterns and analysed the impact of such mutations on software quality. 

Because design anti-patterns negatively affect software quality while design patterns improve it, understanding the evolution of design patterns and design anti-patterns into one another could help developers identify the riskiest design anti-patterns, avoid introducing design anti-patterns, and--or avoid evolving design patterns into such design anti-patterns.

\subsection{Research problem}

In previous works \cite{jaafar2013analysing,jaafar2014anti}, we investigated the static relationships between design anti-patterns and design patterns and how these relationships evolve in time. We studied the relationships between classes playing roles in design patterns and design anti-patterns and reported that static relationships between design patterns and design anti-patterns exist but they are temporary. We also showed that classes playing roles in design anti-patterns and having relationships with design patterns are more change-prone but are less fault-prone than other classes.

In another previous work, \cite{khomh2008design}, we also showed that the design of systems degrades over time, presumably due to the removal (or lack of use) of design patterns and the introduction of design anti-patterns. 

Thus, these studies (and others) considered design patterns and design anti-patterns as unique, atomic entities in each releases of the studied systems. Yet, during software evolution, design (anti)patterns \emph{do} evolve, appearing, disappearing, and mutating into one another. Understanding this evolution of design patterns and design anti-patterns across releases, and in particular their mutations into one another, could help developers avoid mutations that negatively impact software quality while promoting those that improve it.

\subsection{Contributions}

This study is a quasi-replication of a previous study by Jaafar \etal \cite{jaafar2014anti}. Some of the research questions in this paper are similar to those in the previous study \cite{jaafar2014anti}, which pertain to:

\begin{itemize}
\item Computing the probability of design anti-patterns mutations using Markov models.

\item Comparing classes with and without design anti-patterns and their fault-proneness. 
\end{itemize}

 \noindent Moreover, in this study, we consider both design patterns and design anti-pattern mutations during software evolution. We examine the impacts of design patterns and design anti-patterns mutations on change- and fault-proneness. We investigate seven open-source Java software systems to answer to the the following five research contributions:

 \begin{enumerate}
 \item We study how design patterns and design anti-patterns mutate over time using Markov models.

 \item We study the types of changes that occur during design anti-patterns mutations.

 \item We study the impact of these mutations on fault-proneness.

 \item We study the types of changes that lead to design patterns and design anti-patterns mutations.

 \item We study the most fault-prone transitions between design patterns and design anti-patterns.
 \end{enumerate}

We use seven different open-source systems of different sizes and from different application domains: Apache Ignite\footnote{\url{https://ignite.apache.org/}}, Apache Solr \footnote{\url{http://lucene.apache.org/solr/}}, Eclipse IDE \footnote{\url{https://www.eclipse.org/}}, Matsim \footnote{\url{https://matsim.org/}}, Mule \footnote{\url{http://www.mulesoft.org/}}, Nuxeo \footnote{\url{https://www.nuxeo.com/}}, and Ovirt \footnote{\url{https://www.ovirt.org/}}.

We consider thirteen design anti-patterns from \cite{brown1998antipatterns} and eight design patterns \cite{gamma1995design}. For the detection of design anti-patterns and design patterns, we use DECOR \cite{moha2010decor} and DeMIMA \cite{gueheneuc2008demima}. We first detect the occurrences of design anti-patterns and design patterns in all the studied releases of the systems and then investigate the types of mutations: persistent, deleted, introduced, and changed between these snapshots.

Second, we build Markov models \cite{meyn2012markov} to compute the probability values of such mutations. We build one Markov model per studied system. In the models, design anti-patterns and design patterns are nodes while the probabilities of their mutations into one another label the edges between nodes. We compute the probability values by analyzing all the releases of the software systems during the considered period of time.

Third, we use the SZZ algorithm  \cite{sliwerski2005changes} to find fault-inducing commits and investigate the impact of design patterns and--or design anti-patterns mutations on the fault-proneness of classes. 

Fourth, we define thirteen types of change and study them to discover the kinds of changes leading to mutations between design patterns and design anti-patterns. We also study the effects of the change types on fault-proneness.

\subsection{Research Questions} \label{RQuestions}


We use the seven open-source Java software systems to answer the following four research questions: 

\begin{itemize}
\item {\bf RQ1:} \emph{\RQOne} We build Markov models \cite{meyn2012markov} showing which design patterns and--or design anti-patterns mutate into one another during a studied period of evolution. We consider both appearance and disappearance of design patterns and design anti-patterns. We observe that both design patterns and design anti-patterns mutate in the systems. We compute the probabilities of all possible mutations using the Markov models. We also present the most frequent mutations of design patterns and design anti-patterns along the following four mutation types:

\begin{enumerate}
\item Design anti-patterns to some other design anti-patterns;
\item Design anti-patterns to design patterns;
\item Design patterns to design anti-patterns;
\item Design patterns to some other design patterns.
\end{enumerate}

\item {\bf RQ2:} \emph{\RQTwo} Design patterns and design anti-patterns evolve through different types of changes as the system evolves. We define thirteen change types and investigate classes experiencing these change types and participating in design patterns and--or design anti-patterns. We see that different types of changes occur during the evolution of software systems and lead to different mutations. We study the impact of the types of changes on mutations between design patterns and--or design anti-patterns. We present the most prevalent type of changes leading to mutations. 

\item {\bf RQ3:} \emph{\RQThree} Design patterns and design anti-patterns may frequently mutate in other types of patterns during the evolution process. We study whether such mutations are risky regarding fault-proneness. We also present the riskiest transitions among design patterns and design anti-patterns. We observe that classes participating in mutated design anti-patterns are more fault-prone than classes involved in mutated design patterns. We also see that mutations between design anti-patterns and design patterns are more faulty than other mutations. 

\item {\bf RQ4:} \emph{\RQFour} We investigate whether the types of changes to design patterns and design anti-patterns impact fault-proneness. We examine faulty-classes and check whether a mutation occurred during the evolution of these classes. We also examine all changes experienced by the classes during the evolution of the systems. We observe that some of the change types make the systems more fault-prone. We study whether specific types of changes that cause the mutations between design patterns and design anti-patterns are more fault-prone. 
\end{itemize}

The results of these four research questions show that there is a high probability for some design patterns and design anti-patterns to mutate to others types of design patterns and design anti-patterns. The changes that lead to the mutations are mostly structural changes, in particular the addition of large number of attributes or long methods. Results also show that some mutations increase the fault-proneness of the analysed software systems.

\subsection{Organization}

The rest of the paper is organized as follows: Section \ref{sec:Related Works} describes the related work. Section \ref{sec:Methodology} presents the methodology of our study. Section \ref{sec:expSetup} reports its experimental setup. Section \ref{sec:Study Results} presents its results. Section \ref{sec:Discussion} discusses the results and Section \ref{sec:Treats to Validity} threats to their validity. Finally, Section \ref{sec:Conclusion and Future Work} concludes the paper with future work.
\section{Related Work}
\label{sec:Related Works}

\subsection{Design Anti-pattern and Detection}

Webster \etal \cite{webster1995pitfalls} describes design anti-pattern as a solution to a problem that is used frequently but negatively affects software quality. Riel \etal \cite{riel1996object} proposed 61 heuristics of good object-oriented programming that can be used to manually assess a program quality for improving its design and implementation. These heuristics are similar to code smells. Later, Brown \etal \cite{brown1998antipatterns} introduced 40 types of design anti-patterns that form the basis of design anti-patterns detection approaches \cite{van2002java,marinescu2006object,moha2010decor,settas2012enhancing}. 

Several approaches have been proposed to detect design anti-patterns. Van Emden \etal \cite{van2002java} developed JCosmo to visualize the code layout and locate anti-patterns. JCosmo uses primitives and rules to detect design anti-patterns while parsing the source code into an abstract model (similar to the Famix meta-model \cite{tichelaar2000meta}). The goal is to evaluate code quality and help developers to do refactoring. Marinescu \etal \cite{ducasse2004using} combined detection strategies and additional information collected from the documentation of problematic structures in the histories of software systems to improve the detection results.

Settas \etal \cite{settas2012enhancing} proposed Bayesian network-based approach to improve the detection of design anti-patterns. Their approach leverages probabilistic knowledge, which contains the relationships of design anti-patterns regarding their causes, symptoms, and consequences. iPlasma \cite{marinescu2006object} detects design anti-patterns by calculating metrics on C++ or Java source code and by applying some rules that combine the metrics. 

In this paper, we use DECOR to specify and detect design anti-patterns because of its higher detection accuracy \cite{moha2010decor} and wider domain coverage. We present a detailed description of DECOR in Section \ref{ssec:section3.1}.

\subsection{Design Pattern and Detection}

Design patterns in object-oriented software development and their detection have been well-studied in the past two decades \cite{gamma1995design,kramer1996design}. Kramer \etal \cite{kramer1996design} introduced an approach to detect design information directly from C++ header files. Design patterns are represented as Prolog rules that query this design information. Their approach detects five structural design patterns: Adapter, Bridge, Composite, Decorator, and Proxy. 

Voka\v{c} \etal \cite{vokavc2004defect} proposed an approach scoring the similarity between the graph of a design pattern and the graph of a system to identify classes participating in this design pattern. Iacob \etal \cite{iacob2011design} identified proven solutions for recurring design problems using design workshops and system analysis. During design workshops, teams of 3-5 developers designed systems while considering design issues. Then, they analysed a set of systems to recognize how developers should consider design problems in the implementation of existing solutions. 

In this paper, we use DeMIMA to specify and detect design patterns because of its higher detection accuracy \cite{gueheneuc2008demima}, which is described in Section \ref{ssec:section3.2}.

\subsection{Design Anti-pattern and Design Pattern Evolution and their Impact}

Several studies investigated both design anti-pattern and design-pattern evolution. Bieman \etal \cite{bieman2003design} claimed that there is a relative stability in classes participating in design patterns compared to other classes. They showed that large classes are more change-prone than other classes. Voka\v{c} \etal \cite{vokavc2004defect} discussed how different design patterns have different impact on fault-proneness. They studied a large C++ industrial system to prove their claim. 
Gatrell \etal \cite{gatrell2009design} demonstrated that pattern-based classes are more change-prone than other classes. Olbrich \etal \cite{olbrich2009evolution} focused on the historical data of Lucene and Xerces over several years and showed that Blob classes and classes subject to Shotgun Surgery are more change-prone than other classes.

Khomh \etal \cite{khomh2012exploratory} investigated the impact of code smells on the change-proneness of classes. They also studied the influence of design anti-patterns on change- and fault-proneness. They considered 13 design anti-patterns in 54 releases of ArgoUML, Eclipse, Mylyn, and Rhino, and analyzed the probability of classes changing to fix a fault. Their results showed that classes participating in design anti-patterns were significantly more likely to be changed than other classes. This study also investigated two types of changes experienced by classes with design anti-patterns: structural and non-structural changes. Structural changes can modify the class interface while non-structural ones change method bodies. They concluded that structural changes were more likely to occur in classes participating in design anti-patterns.

Yamashita and Moonen \cite{yamashita2013developers} reported that developers cannot fully evaluate the overall maintainability of a software system with code smells alone. They argued that different approaches should be combined to achieve complete and accurate evaluations of software maintainability. Taba \etal \cite{taba2013predicting} claimed that information about design anti-patterns improves the accuracy of fault prediction models. Jaafar \etal \cite{jaafar2013analysing} empirically studied the relationships between design anti-patterns and design patterns. They showed that some design anti-patterns have relationships with design patterns while others do not.
\section{Methodology}
\label{sec:Methodology}

\begin{figure}[ht]
\centering
\includegraphics[scale=0.35]{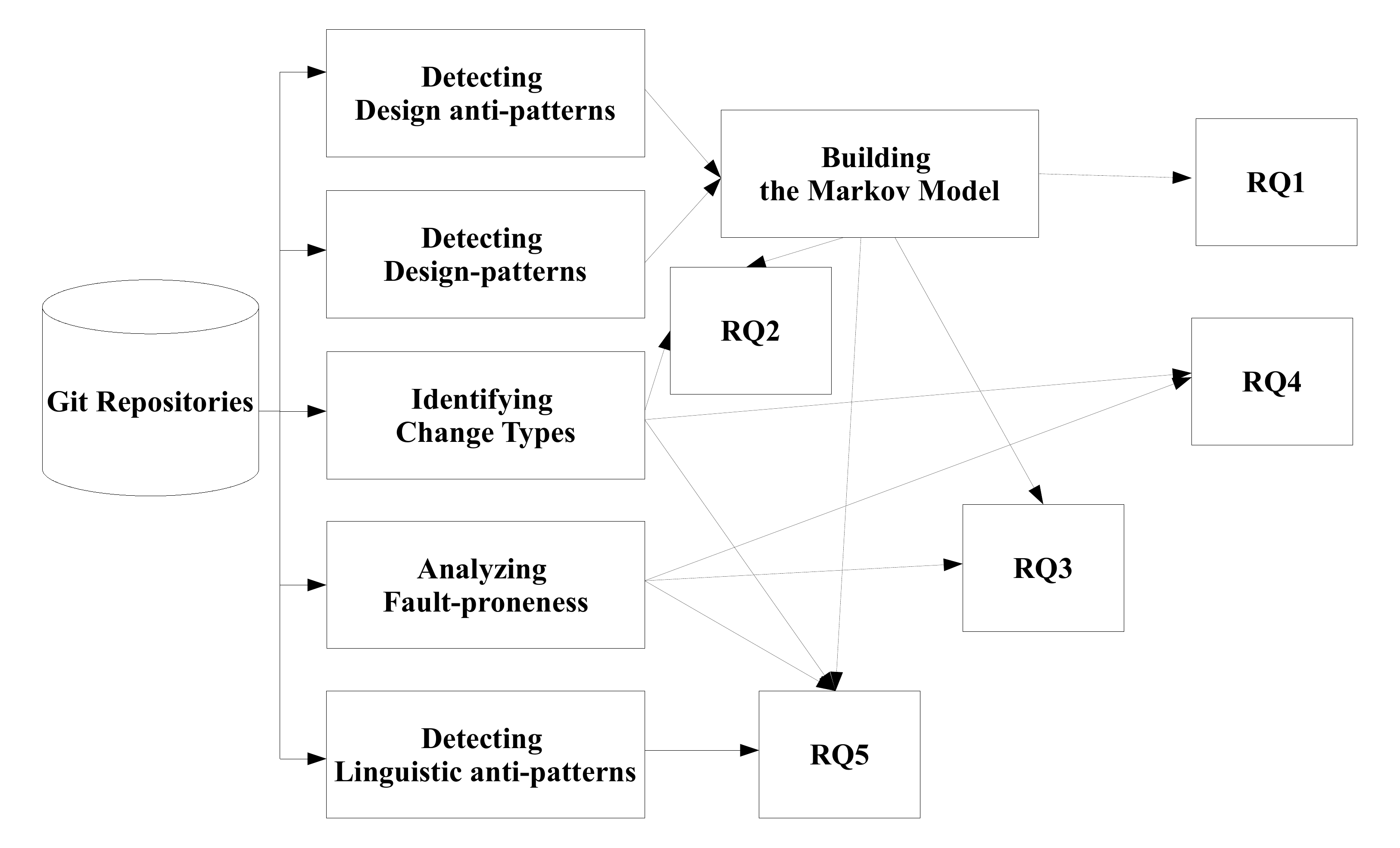}
\caption{Schematic diagram of the methodological steps of the study}
\label{fig:FigureMethod}
\end{figure}

We now describe the general methodology of our study, shown in Figure \ref{fig:FigureMethod}, which includes the main steps of design (anti-)pattern detection, classification of change types, building Markov models, and detection of faulty classes.

We first extract the source code of the studied systems in snapshots taken every 500 commits in their Git repositories. Then, we detect design anti-patterns and design patterns in all the snapshots of the systems. We then create Markov models based on the detected design anti-patterns and design patterns to analyze their behaviors during evolution. We also identify change types and faulty classes throughout the period of evolution that we analyzed. We study all the changed types that lead to fault(s) during evolution. We explain each step in details in the following sub-sections.

\subsection{Detecting Anti-patterns}
\label{ssec:section3.1}

We use Defect DEtection for CORection Approach (DECOR) \cite{moha2010decor} to detect occurrences of design anti-patterns. DECOR offers a domain-specific language to automatically generate design-defect detection algorithms, including design anti-patterns. DECOR uses the Patterns and Abstract-level Description Language meta-model (PADL) \cite{gueheneuc2008demima} and the Primitive, Operators, and Metrics framework (POM) \cite{gueheneuc2004fingerprinting} to detect design anti-patterns in object-oriented systems. The output of DECOR is a list of classes and their roles (if any) in occurrences of design anti-patterns. Moha \etal \cite{moha2010decor} reported that DECOR achieves 100$\%$ recall while having 31$\%$ precision rate in the worst case with an average greater than 60$\%$.

A domain-specific language is more flexible than ad hoc algorithms \cite{moha2010decor} because domain experts (\ie{} developers) can modify the detection rules manually using high-level abstractions, considering the contexts, environments, and characteristics of the analyzed systems. PADL \cite{gueheneuc2008demima} is a meta-model to describe object-oriented systems at different abstraction levels while POM \cite{gueheneuc2004fingerprinting} is a PADL-based framework that implements more than 60 metrics.

\subsection{Detecting Design patterns}
\label{ssec:section3.2}

We use the Design Motif Identification Multilayered Approach (DeMIMA) \cite{gueheneuc2008demima} to detect occurrences of design patterns. DeMIMA traces design motifs (the micro-architecture describing the solutions of the design patterns) in the source code. It discovers idioms relevant to binary class relationships and then provide an idiomatic model of the source code. The model helps to identify design motifs to create a design model of the system. DeMIMA can recover idioms related to both the relationships among classes and design motifs.

DeMIMA uses explanation-based constraint programming to identify occurrences of design motifs using the roles and relationships describing the motifs in PADL models of systems. It reports the micro-architectures that are occurrences of the motifs, including approximations of the motifs. The output of DeMIMA is a list of classes and their roles (if any) in the  occurrences of design patterns. Gu\'{e}h\'{e}neuc and Antoniol \cite{gueheneuc2008demima} report that DeMIMA achieves 100$\%$ recall and 34$\%$ precision.

\subsection{Building Mutation Model}
\label{ssec:section3.4}

We build a Markov model \cite{meyn2012markov} for each studied system to show the mutations between design anti-patterns and design patterns during evolution. For each system, we consider all the patterns whose occurrences we found in its snapshots as nodes in its Markov model. A Markov model shows the set of all possible mutations for one pattern during the evolution of the system. 

First, we obtain two files from two consecutive snapshots of a system, $C_0$ and $C_1$, which contain all the occurrences of design patterns and design anti-patterns in each class of each snapshot. Then, the Markov transition matrix of these two files is a square matrix describing the probabilities of one design anti-pattern and--or design pattern mutating into another. Each row contain the probabilities of mutating from one pattern to all the others. Second, we compare the next two snapshots, $C_1$ and $C_2$, and repeat this algorithm up to snapshots $C_{n-1}$ and $C_n$. Then, we sum up all the mutation probabilities for one given pattern into all the others. We report the averaged summed mutation probabilities divided by the total number of each row. The sum of all the probabilities from any pattern to the others is equal to one. 

\begin{figure*}
\begin{center}
\scalebox{0.8}{
\begin{tikzpicture}[->, >=stealth', auto, semithick, node distance=2.5cm]
\tikzstyle{every state}=[fill=white,draw=black,thick,text=black,scale=1]
\node[state]    (A)                     {$Bu$};
\node[state]    (H)[below right of =A]  {$LM$};
\node[state]    (P)[above right of=A]   {$Cm$};
\node[state]    (Q)[below left of=A]   {$Cp$};
\node[state]    (S)[above left of=A]   {$Sink$};
\node[state]	(V)[left of=A]	        {$De$};

\path
(A) edge[loop right]     node{$0.665$}   (A)
edge        node{$0.003$}       (H)
edge         node{$0.042$}      (P)
edge       node{$0.169$}        (Q)
edge        node{$0.076$}       (V)
edge         node{$0.042$}      (S);

\end{tikzpicture}
}
\caption{Builder (Bu) mutation among the different revisions of Matsim.}
\label{Figure:Bu-Mutations-Matsim2}
\end{center}
\end{figure*}
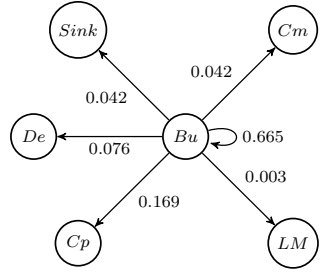

For an example, Figure \ref{Figure:Bu-Mutations-Matsim2} shows a Markov model whose nodes are design anti-patterns and design patterns and edges represent mutations from one pattern to another. Edges are labeled with the probabilities of the mutation from the source patterns to its mutated patterns. This Markov model shows the mutations of the Builder design pattern across the snapshots of Matsim.

\subsection{Analyzing Fault-proneness} \label{ssec:section3.5}

We use the SZZ algorithm \cite{sliwerski2005changes} to identify commits that introduce faults in the systems, \ie{} fault-inducing commits, and thus faulty classes in the system snapshots. For each system, we first apply heuristics \cite{fischer2003populating} to link commits to faults. We use regular expressions to detect fault-IDs in the commit logs. Developers use different conventions for these IDs in their systems so, to ensure accuracy, we tune our heuristics on our dataset incrementally and manually. 

Given a fault $F$ in a system, we extract from its history the files that fixed the fault (fault-fixing files) using \texttt{git diff}. Then, we retrieve the modified and deleted lines from the fault-fixing files. The SZZ algorithm assumes that prior commits that modified these lines are fault-inducing commits. 

To identify such prior commits, for each fault-fixing files, we apply \texttt{git blame} to retrieve a list of previous commits that modified these files. We filter commits whose submission date is later than the fault creation date. We consider the remaining commits as inducing the fault $B$. 

For any $F$, the SZZ algorithm returns a list of commit IDs and fault-inducing files pertaining to $F$. We then use regular expressions to map the object-oriented classes composing the systems with the fault-inducing files.

\subsection{Identifying Change Types}
\label{ssec:section3.6}

Different types of changes can affect software systems with different impacts on fault-proneness. For example, changes to comments are less likely to lead to faults than changes to method invocations. Table \ref{tab:Change_types} shows the change types that we consider, which were also considered in a previous work \cite{an2015empirical}.

\begin{table*}[ht]   
\caption{Change types identified from the source code of the systems studied}
\label{tab:Change_types}
\begin{center}
\scalebox{0.9}{
\begin{tabular}{|l|p{8cm}|} 
\hline
\textbf{Change type} & \textbf{srcML tag(s)}\\  
\hline
	Access & \emph{super, public, private, protected, extern} \\ \hline
 	Class & \emph{extends, class, interface, implements, class\_decl} \\ \hline
   	Code block & \emph{expr\_stmt, expr, block} \\ \hline
  	Comment & \emph{annonation, comment, @type, @format} \\ \hline
  	Control flow & \emph{while, do, if, else, elseif, break, goto, for, foreach, control, continue, then, switch, case, return, incr, default, condition} \\ \hline
 	Declaration & \emph{decl\_stmt, modifier, specifier, decl, function\_decl, literal, label, empty\_stmt, construction\_decl, annonation\_dfn} \\ \hline
 	Exception & \emph{assert, try, catch, throw, throws, finally} \\ \hline
	Import & \emph{import, package} \\ \hline
	Invocation &  \emph{call}\\ \hline
	Method &  \emph{constructor, default, static, type, lambda, function, function\_decl, unit}\\ \hline
	Operator &  \emph{index, synchronized, enum, operator, ternary} \\ \hline
  	Parameter &  \emph{argument, param, parameter\_list, argument\_list, parameter} \\ \hline
    Renaming &  \emph{renaming, name} \\ \hline
\end{tabular}
}
\end{center}
\end{table*} 

We use srcML \cite{srcml} to transform each file in the snapshots of a system into an XML document, in which code elements are tagged by type or function, \eg{} a class declaration, a parameter list, or a control flow statement. Then, we compare the srcML tags between each two subsequent snapshots and extract their differences. The removed tags from the older snapshot and the added tags in the newer snapshot are \emph{changed tags}. We manually group the unique changed tags into a series of \emph{change types}. Table \ref{tab:Change_types} shows the change types and their corresponding srcML (changed) tags. We group some change types together, which have similar impacts on source code. Changes in a same group are likely to have similar impacts on fault-proneness. 

For a given file $F$, in a specific snapshot $S$, our approach yields a list of change types listed in Table \ref{tab:Change_types}. Because we study design (anti-)patterns from commits; in each selected snapshots, we aggregate the change types related to $F$ in the commits \{$C_1$, $C_2$, ..., $C_n$\}, which form that snapshot. 
\section{Experimental Setup}
\label{sec:expSetup}

We now describe the systems and patterns making our experimental setup.

\subsection{Subject Systems} 
\label{ssec:section4.1}

We consider seven Java-based open-source systems for our study, summarised in Table \ref{tab:StudySystems}. We select these systems based on diversity in code size, application domains, and evolution histories. The number of lines of code of these systems range from hundred of thousand to several millions. They belong to different domains, from IDE to database. Some were used in previous studies, which allows some comparisons. These systems have evolved over the years and have many commits/versions to provide a dataset for analyzing pattern mutations and fault-proneness.

However, the choice of systems is inherently a threat to the conclusion and generalisability of any empirical study, which we acknowledge in Section \ref{sec:Treats to Validity}. 

We now briefly describe the subject systems.

\begin{table*} [ht]
\centering
\caption{Analyzed systems}
\scalebox{0.8}{
\begin{tabular}{|l|l|l|r|r|r|}
\hline
\textbf{System} & \textbf{Applicaion domain} &\textbf{\# Commits} & \textbf{LOC} & \textbf{Issue Tracker}\\
\hline \hline
Eclipse for Java & IDE& 281,396 & 9,064,794 & Bugzilla \\
\hline
Nuxeo Platform& Colaboration management& 265,380 & 5,741,131 & Jira \\
\hline
oVirt & Visualization platform & 149,128 & 2,764,655 & Bugzilla \\
\hline
Matsim & Transportation management & 44,200 & 1,602,877 & Atlassian \\
\hline
Apache Solr& Search server& 30,995 & 658,711 & Jira\\
\hline
Apache Ignite&Distributed DB platform& 24,104 & 1,471,036 & Jira\\
\hline
Mule Community Edition&Integration platform & 22,891 & 309,616 & Jira \\
\hline
\end{tabular}
}
\label{tab:StudySystems}
\end{table*}

\noindent\textbf{Eclipse IDE for Java} is an IDE for Java developers. The IDE offers the Java Development Tools (JDT) to develop Java systems. It contains also CVS, SVN, and Git clients. It also includes an XML editor, Mylyn as a task management system, build supports for Maven and WindowBuilder, etc.

\noindent\textbf{Nuxeo}, also called Nuxeo Platform, is an open-source context management and collaboration platform, which provides different information management solutions for developers to build business applications.

\noindent\textbf{oVirt} is a visualization management platform in Java. It provides a centralized management of resources, storage, and virtual machines, which allows managing enterprise infrastructure.

\noindent\textbf{Matism} is a framework to build large-scale transport simulations. Its development team provides a comprehensive documentation for users and developers to ease usability and maintainability. 

\noindent\textbf{Apache Solr} from the Apache Lucene project is an open-source Java search server for Web sites, databases, and files. It is popular and fast, using Lucene Java search library at its core. It runs as a standalone full-text search server. 

\noindent\textbf{Apache Ignite} is a in-memory computing platform used as database and caching system. It helps solving problems related to speed and scalability and can be used to speed up relational and NoSQL databases.

\noindent\textbf{Mule} is the run-time engine of a Java-based enterprise service bus (ESB) and integration platform. Developers can connect applications quickly and easily to exchange data. It allows service creation and hosting, service mediation, message routing, and data transformation.

\subsection{Analyzed Patterns}

\subsubsection{Anti-patterns}

We select thirteen anti-patterns in our study. These anti-patterns introduced by Brown \etal \cite{brown1998antipatterns} express problems with data, complexity, size, and the features related to classes \cite{khomh2012exploratory}. They have been studied in previous work \cite{khomh2012exploratory}. We summarize their definitions below, details are available elsewhere \cite{romano2012analyzing}:

\begin{itemize}
\item AntiSingleton (AS): A class that provides mutable class variables, which could be used as global variables.

\item Blob (Bl) or God Class (GC): A class that is too large and not cohesive enough, which monopolizes most of the system processing, takes most of the decisions, and is associated to data classes.

\item ClassDataShouldBePrivate (CS): A class that exposes its fields, thus violating the principle of encapsulation.

\item ComplexClass (CC): A class that has (at least) one large method and complex method, in terms of cyclomatic complexity and line of codes LOCs.

\item LargeClass (LC): A class that has (at least) one large method, in terms of LOCs.

\item LazyClass (LZC): A class that has few fields and methods that are complex.

\item LongMethod (LM): A class that has (at least) one method that is overly long, in terms of LOCs.

\item LongParameterList (LP): A class that has (at least) one method with a long list of parameters with respect to the average numbers of parameters per methods.

\item MessageChain (MCh): A class that uses a long chain of method invocations to realize one of its functionality.

\item RefusedParentBequest (RP): A class that overrides methods using empty bodies.

\item SpaghettiCode (SC): A class declaring long methods which do not have any parameters. These methods are complex, with a complicated control flow. The class does not use polymorphism and--or inheritance.

\item SpeculativeGenerality (SG): A class that is defined as abstract but that has very few children, which do not make use of its methods.

\item SwissArmyKnife (SA): A class whose methods can be divided into disjoint sets, providing different, unrelated functionalities.
\end{itemize}

\subsubsection{Design Patterns}

We consider eight design patterns presented in Gamma \etal \cite{gamma1995design}, which we select due to their popularity and because previous works also studied them \cite{tsantalis2006design,vlissides1995design}. Their complete definitions and specifications are available in \cite{vlissides1995design,khomh2009playing}:

\begin{itemize}
\item Builder (Bu): A pattern to separate the construction of a complex object from its representation.

\item Command (Cm): A pattern to encapsulate a request as an object.

\item Composite (Cp): A pattern that composes objects into tree structures to represent part-whole hierarchies. Composite lets clients treat individual objects and compositions of objects uniformly.

\item Decorator (De): A pattern that attaches additional responsibilities to an object dynamically. Decorator provides a flexible alternative to sub-classing for extending functionality.

\item Factory Method (FM): A pattern that defines an API for object creation in which subclasses choose the class to instantiate.

\item Observer (Ob): A pattern that defines a one-to-many dependency between objects to notify all the object dependent on one object when it changes.

\item Prototype (Pt): A pattern that specifies the kind of objects to create using a prototypical instance.

\item Singleton (Si): A pattern that restricts the instantiation of a class to one object to coordinate actions across a system.
\end{itemize}

\section{Results and Analysis}
\label{sec:Study Results}

We now present the results of our study and answer our five research questions. 

\subsection{\textbf{RQ1:} \textit{\RQOne}}

\paragraph{\textbf{Motivation}} Understanding the evolution of patterns is important because it can help developers to identify and circumvent risky design patterns and prevent the appearance of design anti-patterns \cite{jaafar2014anti}. While some tools can find software entities and their evolution patterns automatically, \eg{} \cite{van2002java,lanza2007object,rapu2004using,vaucher2009tracking}, no previous work investigated the mutation of design (anti-)patterns.

\paragraph{\textbf{Computing probability values for all possible mutations}} We apply the detection tools described in Section \ref{sec:Methodology} on snapshots of each of the systems listed in Table \ref{tab:StudySystems}. Each snapshot contains a large number of classes, which may participate in different types of design anti-patterns and--or design patterns. 

We take snapshots every 500 commits in the evolution histories of the systems. This commit interval period is adequate to detect changes occurring between two subsequent snapshots \cite{hassan2009predicting,canfora2010exploratory}.

We automatically compare each two subsequent snapshots to compute the numbers of added or deleted occurrences of design patterns and design anti-patterns. We build one Markov model for each system to show the probabilities of mutations between design patterns and design anti-patterns.

\begin{landscape}
\begin{table*}
\caption{Change probabilities of design anti-patterns and design patterns in Eclipse IDE}
\setlength\tabcolsep{0.08cm}
\scriptsize
\centering
\scalebox{0.73}{
{\renewcommand{\arraystretch}{1.05}
\begin{tabular}
{|l|l||l|l|l|l|l|l|l|l|l|l|l|l|l||l|l|l|l|l|l|l|l||l|p{0.7cm}|p{0.9cm}}
\cline{2-24}
\multicolumn{1}{l|}{}& Source & AS & Bl & CS & CC & LC & LZC & LM & LP & MCh & RP & SC & SG & SA & Bu & Cm & Cp & FM & De & Ob & PT & Si & Sink \\
\hline
AS & 0.013 & \cellcolor[gray]{0.8}{\textbf{0.972}} & 0.000 & 0.000 & 0.000 & 0.000 & 0.000 & 0.000 & 0.000 & 0.000 & 0.000 & 0.000 & 0.000 & 0.000 & 0.000 & 0.000 & 0.000 & 0.000 & 0.000 & 0.000 & 0.000 & 0.000 & 0.015 \\
\hline
Bl & 0.007 & \cellcolor[gray]{0.8}{\textbf{0.27}} & \cellcolor[gray]{0.8}{\textbf{0.447}} & 0.000 & 0.000 & 0.000 & 0.000 & 0.000 & 0.000 & 0.000 & 0.000 & 0.000 & 0.000 & 0.000 & 0.000 & 0.000 & 0.000 & 0.000 & 0.000 & 0.000 & 0.000 & 0.000 & \cellcolor[gray]{0.8}{\textbf{0.277}} \\
\hline
CS & 0.001 & 0.000 & 0.001 & \cellcolor[gray]{0.8}{\textbf{0.993}} & 0.000 & 0.000 & 0.000 & 0.000 & 0.000 & 0.000 & 0.000 & 0.000 & 0.000 & 0.000 & 0.000 & 0.000 & 0.000 & 0.000 & 0.000 & 0.000 & 0.000 & 0.000 & 0.005 \\
\hline
CC & 0.000 & 0.000 & 0.000 & 0.012 & \cellcolor[gray]{0.8}{\textbf{0.973}} & 0.000 & 0.000 & 0.000 & 0.000 & 0.000 & 0.000 & 0.000 & 0.000 & 0.000 & 0.000 & 0.000 & 0.000 & 0.000 & 0.000 & 0.000 & 0.000 & 0.000 & 0.014 \\
\hline
LC & 0.000 & 0.000 & 0.000 & 0.000 & \cellcolor[gray]{0.8}{\textbf{0.5}} & \cellcolor[gray]{0.8}{\textbf{0.5}} & 0.000 & 0.000 & 0.000 & 0.000 & 0.000 & 0.000 & 0.000 & 0.000 & 0.000 & 0.000 & 0.000 & 0.000 & 0.000 & 0.000 & 0.000 & 0.000 & 0.000 \\
\hline
LZC & 0.000 & 0.000 & 0.000 & 0.000 & 0.000 & 0.012 & \cellcolor[gray]{0.8}{\textbf{0.982}} & 0.000 & 0.000 & 0.000 & 0.000 & 0.000 & 0.000 & 0.000 & 0.000 & 0.000 & 0.000 & 0.000 & 0.000 & 0.000 & 0.000 & 0.000 & 0.006 \\
\hline
LM & 0.000 & 0.000 & 0.000 & 0.001 & 0.000 & 0.000 & 0.024 & \cellcolor[gray]{0.8}{\textbf{0.95}} & 0.000 & 0.000 & 0.000 & 0.000 & 0.000 & 0.000 & 0.000 & 0.000 & 0.000 & 0.000 & 0.000 & 0.000 & 0.000 & 0.000 & 0.025 \\
\hline
\end{tabular}
}
}
\label{tab:EclipseMarkov}
\end{table*}

\begin{table*} 
\caption{Change probabilities of design anti-patterns and design patterns in Nuxeo}
\setlength\tabcolsep{0.08cm}
\scriptsize
\centering
\scalebox{0.73}{
{\renewcommand{\arraystretch}{1.05}
%
}
\caption{FactoryMethod (FM) mutation in Eclipse.}
\label{Figure:FM-Mutations-Eclipse}
\end{center}
\end{figure*}

\begin{figure}
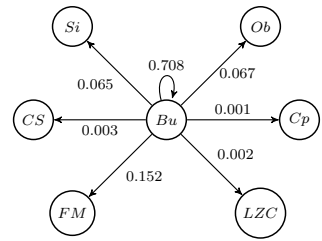

\begin{center}
\scalebox{0.7}{
%
}
\caption{Builder (Bu) mutation in Nuxeo.}
\label{Figure:Bu-Mutations-Nuxeo}
\end{center}
\end{figure} 

\begin{figure}
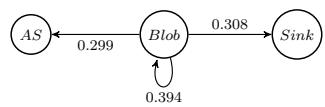

\begin{center}
\scalebox{0.7}{
%
}
\caption{Blob (Bl) mutation in oVirt.}
\label{Figure:Bl-Mutations-Ovirt}
\end{center}
\end{figure}

\begin{figure*}
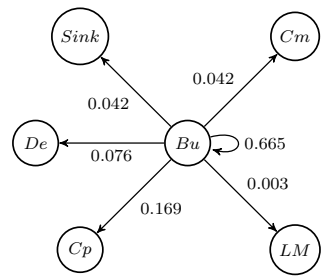

\begin{center}
\scalebox{0.8}{
%
}
\caption{Builder (Bu) mutation among the different snapshots of Matsim.}
\label{Figure:Bu-Mutations-Matsim}
\end{center}
\end{figure*}

\begin{landscape}
\begin{table*}
\caption{Change probabilities of design anti-patterns and design patterns in oVirt}
\setlength\tabcolsep{0.08cm}
\scriptsize
\centering
\scalebox{0.73}{
{\renewcommand{\arraystretch}{1.05}
\begin{tabular}
{|l|l||l|l|l|l|l|l|l|l|l|l|l|l|l||l|l|l|l|l|l|l|l||p{0.7cm}|p{0.9cm}}
\cline{2-24}
\multicolumn{1}{l|}{}& Source & AS & Bl & CS & CC & LC & LZC & LM & LP & MCh & RP & SC & SG & SA & Bu & Cm & Cp & FM & De & Ob & PT & Si & Sink \\
\hline
AS & 0.018 & \cellcolor[gray]{0.8}{\textbf{0.971}} & 0.000 & 0.000 & 0.000 & 0.000 & 0.000 & 0.000 & 0.000 & 0.000 & 0.000 & 0.000 & 0.000 & 0.000 & 0.000 & 0.000 & 0.000 & 0.000 & 0.000 & 0.000 & 0.000 & 0.000 & 0.011\\
\hline
Bl & 0.000 & \cellcolor[gray]{0.8}{\textbf{0.299}} & \cellcolor[gray]{0.8}{\textbf{0.394}} & 0.000 & 0.000 & 0.000 & 0.000 & 0.000 & 0.000 & 0.000 & 0.000 & 0.000 & 0.000 & 0.000 & 0.000 & 0.000 & 0.000 & 0.000 & 0.000 & 0.000 & 0.000 & 0.000 & \cellcolor[gray]{0.8}{\textbf{0.308}}\\
\hline
CS & 0.006 & 0.000 & 0.003 & \cellcolor[gray]{0.8}{\textbf{0.982}} & 0.000 & 0.000 & 0.000 & 0.000 & 0.000 & 0.000 & 0.000 & 0.000 & 0.000 & 0.000 & 0.000 & 0.000 & 0.000 & 0.000 & 0.000 & 0.000 & 0.000 & 0.000 & 0.009 \\
\hline
CC & 0.000 & 0.000 & 0.000 & 0.012 & \cellcolor[gray]{0.8}{\textbf{0.975}} & 0.000 & 0.000 & 0.000 & 0.000 & 0.000 & 0.000 & 0.000 & 0.000 & 0.000 & 0.000 & 0.000 & 0.000 & 0.000 & 0.000 & 0.000 & 0.000 & 0.000 & 0.013 \\
\hline
LC & 0.000 & 0.000 & 0.000 & 0.000 & 0.000 & \cellcolor[gray]{0.8}{\textbf{1}} & 0.000 & 0.000 & 0.000 & 0.000 & 0.000 & 0.000 & 0.000 & 0.000 & 0.000 & 0.000 & 0.000 & 0.000 & 0.000 & 0.000 & 0.000 & 0.000 & 0.000 \\
\hline
LZC & 0.000 & 0.000 & 0.000 & 0.000 & 0.000 & 0.001 & \cellcolor[gray]{0.8}{\textbf{0.998}} & 0.000 & 0.000 & 0.000 & 0.000 & 0.000 & 0.000 & 0.000 & 0.000 & 0.000 & 0.000 & 0.000 & 0.000 & 0.000 & 0.000 & 0.000 & 0.001 \\
\hline
LM & 0.000 & 0.000 & 0.000 & 0.000 & 0.000 & 0.000 & 0.012 & \cellcolor[gray]{0.8}{\textbf{0.977}} & 0.000 & 0.000 & 0.000 & 0.000 & 0.000 & 0.000 & 0.000 & 0.000 & 0.000 & 0.000 & 0.000 & 0.000 & 0.000 & 0.000 & 0.011 \\
\hline
\end{tabular}
}}
\label{tab:Ovirtmarkov}
\end{table*}

\begin{table*} 
\caption{Change probabilities of design anti-patterns and design patterns in Matsim}
\setlength\tabcolsep{0.08cm}
\scriptsize
\centering
\scalebox{0.73}{
{\renewcommand{\arraystretch}{1.05}
\begin{tabular}
{|l|l||l|l|l|l|l|l|l|l|l|l|l|l|l||l|l|l|l|l|l|l|l||p{0.7cm}|p{0.9cm}}
\cline{2-24}
\multicolumn{1}{l|}{}& Source & AS & Bl & CS & CC & LC & LZC & LM & LP & MCh & RP & SC & SG & SA & Bu & Cm & Cp & FM & De & Ob & PT & Si & Sink \\
\hline
AS & 0.066 & \cellcolor[gray]{0.8}{\textbf{0.893}} & 0.000 & 0.000 & 0.000 & 0.000 & 0.000 & 0.000 & 0.000 & 0.000 & 0.000 & 0.000 & 0.000 & 0.000 & 0.000 & 0.000 & 0.000 & 0.040 & 0.000 & 0.000 & 0.000 & 0.000 & 0.000\\
\hline
Bl & 0.003 & \cellcolor[gray]{0.8}{\textbf{0.372}} & \cellcolor[gray]{0.8}{\textbf{0.279}} & 0.000 & 0.000 & 0.000 & 0.000 & 0.000 & 0.000 & 0.000 & 0.000 & 0.000 & 0.000 & 0.000 & 0.000 & 0.000 & 0.000 & \cellcolor[gray]{0.8}{\textbf{0.346}} & 0.000 & 0.000 & 0.000 & 0.000 & 0.000\\
\hline
CS & 0.025 & 0.000 & 0.037 & \cellcolor[gray]{0.8}{\textbf{0.9}} & 0.000 & 0.000 & 0.000 & 0.000 & 0.000 & 0.000 & 0.000 & 0.000 & 0.000 & 0.000 & 0.000 & 0.000 & 0.000 & 0.038 & 0.000 & 0.000 & 0.000 & 0.000 & 0.000 \\
\hline
CC & 0.004 & 0.001 & 0.002 & 0.075 & \cellcolor[gray]{0.8}{\textbf{0.848}} & 0.000 & 0.000 & 0.000 & 0.000 & 0.000 & 0.000 & 0.000 & 0.000 & 0.000 & 0.000 & 0.000 & 0.000 & 0.069 & 0.000 & 0.000 & 0.000 & 0.000 & 0.000 \\
\hline
LC & 0.000 & 0.000 & 0.000 & 0.000 & 0.000 & \cellcolor[gray]{0.8}{\textbf{1}} & 0.000 & 0.000 & 0.000 & 0.000 & 0.000 & 0.000 & 0.000 & 0.000 & 0.000 & 0.000 & 0.000 & 0.000 & 0.000 & 0.000 & 0.000 & 0.000 & 0.000 \\
\hline
LZC & 0.000 & 0.003 & 0.000 & 0.000 & 0.000 & \cellcolor[gray]{0.8}{\textbf{0.1}} & \cellcolor[gray]{0.8}{\textbf{0.831}} & 0.000 & 0.000 & 0.000 & 0.000 & 0.000 & 0.000 & 0.000 & 0.000 & 0.000 & 0.000 & 0.066 & 0.000 & 0.000 & 0.000 & 0.000 & 0.000 \\
\hline
LM & 0.003 & 0.001 & 0.002 & 0.013 & 0.000 & 0.000 & 0.063 & \cellcolor[gray]{0.8}{\textbf{0.86}} & 0.000 & 0.000 & 0.000 & 0.000 & 0.000 & 0.000 & 0.000 & 0.000 & 0.000 & 0.059 & 0.000 & 0.000 & 0.000 & 0.000 & 0.000 \\
\hline
\end{tabular}
}}
\label{tab:MatsimMarkov}
\end{table*}
\end{landscape}



\begin{landscape}
\begin{table*}
\caption{Change probabilities of design anti-patterns and design patterns in ApacheSolr}
\setlength\tabcolsep{0.08cm}
\scriptsize
\centering
\scalebox{0.73}{
{\renewcommand{\arraystretch}{1.05}
\begin{tabular}
{|l|l||l|l|l|l|l|l|l|l|l|l|l|l|l||l|l|l|l|l|l|l|l||p{0.9cm}|p{0.9cm}}
\cline{2-24}
\multicolumn{1}{l|}{}& Source & AS & Bl & CS & CC & LC & LZC & LM & LP & MCh & RP & SC & SG & SA & Bu & Cm & Cp & FM & De & Ob & PT & Si & Sink \\
\hline
AS & 0.000 & \cellcolor[gray]{0.8}{\textbf{1}} & 0.000 & 0.000 & 0.000 & 0.000 & 0.000 & 0.000 & 0.000 & 0.000 & 0.000 & 0.000 & 0.000 & 0.000 & 0.000 & 0.000 & 0.000 & 0.000 & 0.000 & 0.000 & 0.000 & 0.000 & 0.000 \\
\hline
Bl & 0.000 & \cellcolor[gray]{0.8}{\textbf{0.321}} & \cellcolor[gray]{0.8}{\textbf{0.365}} & 0.000 & 0.000 & 0.000 & 0.000 & 0.000 & 0.000 & 0.000 & 0.000 & 0.000 & 0.000 & 0.000 & 0.000 & 0.000 & 0.000 & 0.000 & 0.000 & 0.000 & 0.000 & 0.000 & \cellcolor[gray]{0.8}{\textbf{0.313}} \\
\hline
CS & 0.000 & 0.000 & 0.009 & \cellcolor[gray]{0.8}{\textbf{0.983}} & 0.000 & 0.000 & 0.000 & 0.000 & 0.000 & 0.000 & 0.000 & 0.000 & 0.000 & 0.000 & 0.000 & 0.000 & 0.000 & 0.000 & 0.000 & 0.000 & 0.000 & 0.000 & 0.008 \\
\hline
CC & 0.000 & 0.000 & 0.000 & 0.033 & \cellcolor[gray]{0.8}{\textbf{0.93}} & 0.000 & 0.000 & 0.000 & 0.000 & 0.000 & 0.000 & 0.000 & 0.000 & 0.000 & 0.000 & 0.000 & 0.000 & 0.000 & 0.000 & 0.000 & 0.000 & 0.000 & 0.037 \\
\hline
LC & 0.000 & 0.000 & 0.000 & 0.000 & 0.000 & \cellcolor[gray]{0.8}{\textbf{1}} & 0.000 & 0.000 & 0.000 & 0.000 & 0.000 & 0.000 & 0.000 & 0.000 & 0.000 & 0.000 & 0.000 & 0.000 & 0.000 & 0.000 & 0.000 & 0.000 & 0.000 \\
\hline
LZC & 0.000 & 0.000 & 0.000 & 0.000 & 0.000 & 0.072 & \cellcolor[gray]{0.8}{\textbf{0.859}} & 0.000 & 0.000 & 0.000 & 0.000 & 0.000 & 0.000 & 0.000 & 0.000 & 0.000 & 0.000 & 0.000 & 0.000 & 0.000 & 0.000 & 0.000 & 0.069 \\
\hline
LM & 0.000 & 0.000 & 0.001 & 0.002 & 0.000 & 0.000 & 0.033 & \cellcolor[gray]{0.8}{\textbf{0.928}} & 0.000 & 0.000 & 0.000 & 0.000 & 0.000 & 0.000 & 0.000 & 0.000 & 0.000 & 0.000 & 0.000 & 0.000 & 0.000 & 0.000 & 0.036 \\
\hline
\end{tabular}
}
}
\label{tab:SolrMarkov}
\end{table*}

\begin{table*} 
\caption{Change probabilities of design anti-patterns and design patterns in ApacheIgnite}
\setlength\tabcolsep{0.08cm}
\scriptsize
\centering
\scalebox{0.735}{
{\renewcommand{\arraystretch}{1.05}
\begin{tabular}
{|l|l||l|l|l|l|l|l|l|l|l|l|l|l|l||l|l|l|l|l|l|l|l||p{0.9cm}|p{0.9cm}}
\cline{2-24}
\multicolumn{1}{l|}{}& Source & AS & Bl & CS & CC & LC & LZC & LM & LP & MCh & RP & SC & SG & SA & Bu & Cm & Cp & FM & De & Ob & PT & Si & Sink \\
\hline
AS & 0.000 & \cellcolor[gray]{0.8}{\textbf{1}} & 0.000 & 0.000 & 0.000 & 0.000 & 0.000 & 0.000 & 0.000 & 0.000 & 0.000 & 0.000 & 0.000 & 0.000 & 0.000 & 0.000 & 0.000 & 0.000 & 0.000 & 0.000 & 0.000 & 0.000 & 0.000 \\
\hline
Bl & 0.000 & \cellcolor[gray]{0.8}{\textbf{0.375}} & \cellcolor[gray]{0.8}{\textbf{0.33}} & 0.000 & 0.000 & 0.000 & 0.000 & 0.000 & 0.000 & 0.000 & 0.000 & 0.000 & 0.000 & 0.000 & 0.000 & 0.000 & 0.000 & 0.000 & 0.000 & 0.000 & 0.000 & 0.000 & \cellcolor[gray]{0.8}{\textbf{0.295}} \\
\hline
CS & 0.000 & 0.000 & 0.000 & \cellcolor[gray]{0.8}{\textbf{1}} & 0.000 & 0.000 & 0.000 & 0.000 & 0.000 & 0.000 & 0.000 & 0.000 &  0.000 & 0.000 & 0.000 & 0.000 & 0.000 & 0.000 & 0.000 & 0.000 & 0.000 & 0.000 & 0.000 \\
\hline
CC & 0.000 & 0.000 & 0.000 & 0.053 & \cellcolor[gray]{0.8}{\textbf{0.905}} & 0.000 & 0.000 & 0.000 & 0.000 & 0.000 & 0.000 & 0.000 & 0.000 & 0.000 & 0.000 & 0.000 & 0.000 & 0.000 & 0.000 & 0.000 & 0.000 & 0.000 & 0.042 \\
\hline
LC & 0.000 & 0.000 & 0.000 & 0.000 & 0.000 & \cellcolor[gray]{0.8}{\textbf{1}} & 0.000 & 0.000 & 0.000 & 0.000 & 0.000 & 0.000 & 0.000 & 0.000 & 0.000 & 0.000 & 0.000 & 0.000 & 0.000 & 0.000 & 0.000 & 0.000 & 0.000 \\
\hline
LZC & 0.000 & 0.000 & 0.000 & 0.000 & 0.000 & 0.015 & \cellcolor[gray]{0.8}{\textbf{0.97}} & 0.000 & 0.000 & 0.000 & 0.000 & 0.000 & 0.000 & 0.000 & 0.000 & 0.000 & 0.000 & 0.000 & 0.000 & 0.000 & 0.000 & 0.000 & 0.015 \\
\hline
LM & 0.000 & 0.000 & 0.000 & 0.004 & 0.000 & 0.000 & 0.045 & \cellcolor[gray]{0.8}{\textbf{0.905}} & 0.000 & 0.000 & 0.000 & 0.000 & 0.000 & 0.000 & 0.000 & 0.000 & 0.000 & 0.000 & 0.000 & 0.000 & 0.000 & 0.000 & 0.047 \\
\hline
\end{tabular}
}}
\label{tab:IgniteMarkov}
\end{table*}

\end{landscape}

\begin{landscape}
\begin{table*}
\caption{Change probabilities of design anti-patterns and design patterns in Mule}
\setlength\tabcolsep{0.08cm}
\scriptsize
\centering
\scalebox{0.73}{
{\renewcommand{\arraystretch}{1.05}
\begin{tabular}
{|l|l||l|l|l|l|l|l|l|l|l|l|l|l|l||l|l|l|l|l|l|l|l||p{0.7cm}|p{0.9cm}}
\cline{2-24}
\multicolumn{1}{l|}{}& Source & AS & Bl & CS & CC & LC & LZC & LM & LP & MCh & RP & SC & SG & SA & Bu & Cm & Cp & FM & De & Ob & PT & Si & Sink \\
\hline
AS & 0.032 & \cellcolor[gray]{0.8}{\textbf{0.937}} & 0.000 & 0.000 & 0.000 & 0.000 & 0.000 & 0.000 & 0.000 & 0.000 & 0.000 & 0.000 & 0.000 & 0.000 & 0.000 & 0.000 & 0.000 & 0.030 & 0.000 & 0.000 & 0.000 & 0.000 & 0.000\\
\hline
Bl & 0.000 & \cellcolor[gray]{0.8}{\textbf{0.313}} & \cellcolor[gray]{0.8}{\textbf{0.313}} & 0.000 & 0.000 & 0.000 & 0.000 & 0.000 & 0.000 & 0.000 & 0.000 & 0.000 & 0.000 & 0.000 & 0.000 & 0.000 & 0.000 & \cellcolor[gray]{0.8}{\textbf{0.374}} & 0.000 & 0.000 & 0.000 & 0.000 & 0.000\\
\hline
CS & 0.003 & 0.000 & 0.006 & \cellcolor[gray]{0.8}{\textbf{0.963}} & 0.000 & 0.000 & 0.000 & 0.000 & 0.000 & 0.000 & 0.000 & 0.000 & 0.000 & 0.000 & 0.000 & 0.000 & 0.000 & 0.028 & 0.000 & 0.000 & 0.000 & 0.000 & 0.000 \\
\hline
CC & 0.001 & 0.000 & 0.000 & 0.071 & \cellcolor[gray]{0.8}{\textbf{0.843}} & 0.000 & 0.000 & 0.000 & 0.000 & 0.000 & 0.000 & 0.000 & 0.000 & 0.000 & 0.000 & 0.000 & 0.000 & 0.084 & 0.000 & 0.000 & 0.000 & 0.000 & 0.000 \\
\hline
LC & 0.000 & 0.000 & 0.000 & 0.000 & 0.000 & \cellcolor[gray]{0.8}{\textbf{1}} & 0.000 & 0.000 & 0.000 & 0.000 & 0.000 & 0.000 & 0.000 & 0.000 & 0.000 & 0.000 & 0.000 & 0.000 & 0.000 & 0.000 & 0.000 & 0.000 & 0.000 \\
\hline
LZC & 0.000 & 0.000 & 0.000 & 0.000 & 0.000 & 0.030 & \cellcolor[gray]{0.8}{\textbf{0.946}} & 0.000 & 0.000 & 0.000 & 0.000 & 0.000 & 0.000 & 0.000 & 0.000 & 0.000 & 0.000 & 0.024 & 0.000 & 0.000 & 0.000 & 0.000 & 0.000 \\
\hline
LM & 0.000 & 0.000 & 0.000 & 0.010 & 0.000 & 0.000 & 0.039 & \cellcolor[gray]{0.8}{\textbf{0.907}} & 0.000 & 0.000 & 0.000 & 0.000 & 0.000 & 0.000 & 0.000 & 0.000 & 0.000 & 0.044 & 0.000 & 0.000 & 0.000 & 0.000 & 0.000 \\
\hline
\end{tabular}
}}
\label{tab:Mulemarkov}
\end{table*}

\begin{table*}
\caption{Change probabilities of design anti-patterns and design patterns in all studied systems}
\setlength\tabcolsep{0.08cm}
\scriptsize
\centering
\scalebox{0.73}{
{\renewcommand{\arraystretch}{1.05}
%
}
\caption{LongParameterList (LP) mutation in ApacheSolr.}
\label{Figure:LP-Mutations-ApacheSolr}
\end{center}
\end{figure*}

\begin{figure*}
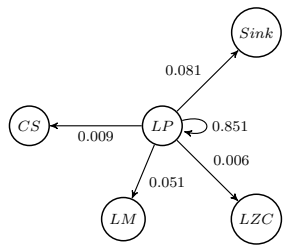
 
\begin{center}
\scalebox{0.7}{
%
}
\caption{LongParameterList (LP) mutation in ApacheIgnite.}
\label{Figure:LP-Mutations-ApacheIgnite}
\end{center}
\end{figure*}

\begin{figure}
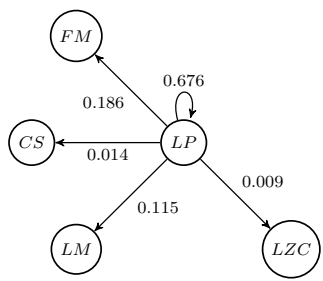

\begin{center}
\scalebox{0.8}{
%
}
\caption{LongParameterList (LP) mutation in Mule.}
\label{Figure:LP-Mutations-Mule}
\end{center}
\end{figure}

Tables \ref{tab:EclipseMarkov} to \ref{tab:Mulemarkov} show the mutations between design patterns and design anti-patterns that occurred in their evolutions. Table \ref{tab:AllSystems} aggregates all the mutations in all the systems. We added two additional states (source and sink) to describe the appearances of design (anti-)patterns (sources) and the disappearance of some design (anti-)patterns (sinks).


\begin{table*} 
\centering
\caption{Standard deviation values and confidence levels}
\scalebox{0.9}{
\renewcommand{\arraystretch}{1.05}
\begin{math}
\begin{tabular}{|l|r|r|r|}
\hline
\textbf{Systems} & \textbf{Standard Deviation} & \textbf{Confidence Level} & \textbf{Margin of Error}\\
 \hline \hline
 Apache Ignite & 0.1966 & 90\%, 1.645 $\sigma x$ & 0.04347$\pm$0.0147 ($\pm$33.87\%)\\ 
 \hline
 Apache Solr & 0.1921 & 90\%, 1.645 $\sigma x$ & 0.04343$\pm$0.0144 ($\pm$33.12\%) \\
 \hline
 Eclipse & 0.1833 & 90\%, 1.645 $\sigma x$ & 0.04317$\pm$0.0137 ($\pm$31.79\%) \\
\hline
 Matsim & 0.1722 & 90\%, 1.645 $\sigma x$ & 0.04304 $\pm$0.0129 ($\pm$29.96\%) \\
\hline
 Mule & 0.1766 & 90\%, 1.645 $\sigma x$ & 0.04345 $\pm$0.0132($\pm$30.43\%) \\
\hline
 Nuxeo & 0.1968 & 90\%, 1.645 $\sigma x$ & 0.0451 $\pm$0.0147 ($\pm$32.67\%) \\
\hline
 oVirt & 0.1961 & 90\%, 1.645 $\sigma x$ & 0.04352 $\pm$0.0147 ($\pm$33.73\%) \\
\hline
\end{tabular}
\end{math}
}
\label{tab:standardDeviation}
\end{table*}

The mutation probabilities shown in previous tables are percentages that may hide outliers. Therefore, we also calculate the standard deviation values among these probabilities. 
We found that the probabilities across snapshots have low standard-deviation values, as shown in Table \ref{tab:standardDeviation}, with the highest value of 0.196 for Nuxeo. 


Table \ref{tab:standardDeviation} shows a systematic analysis of the confidence levels of our results. We computed the standard-deviation values and confidence intervals of our results for a confidence level of 90\% as follows:

\begin{equation}
    \sigma = \sqrt{\frac{1}{N}\sum_{i=1}^{N}{ (x_i-\mu)}^2}
\end{equation}

\noindent where $x_i$ is each value from the population (mutations probabilities), $\mu$ is the mean of the population, and $N$ is the size of the population, \ie{} the total number of mutations in all the snapshots of a system.

With the standard deviation known, we compute the confidence interval for a population mean as:

\begin{equation}
    \bar{X}\pm Z \times {\frac{\sigma}{\sqrt{N}}}
\end{equation}

\noindent where $\bar{X}$ is plus or minus a margin of error, $Z$ is the Z-value for the chosen confidence level, $\sigma$ is a standard deviation and $N$ is the size of the population.

We observe that, for a confidence level of 90\%, the confidence intervals, which indicate how much we can expect the results to reflect the observations from the overall population, were around 30\% in all the analysed systems. We consider these values of confidence levels and intervals acceptable to deduce trends and infer conclusions. Indeed, while we could not find similar discussions and numbers in other software-engineering papers, we observed similar values used to deduce trends in other domains, \eg{} public health \cite{strazzullo2009salt}.

\begin{table*} 
\centering
\caption{Mean values of the mutations of design anti-patterns and design pattern occurrences mutated in all the snapshots of each system}
\scalebox{0.8}{
\renewcommand{\arraystretch}{1.1}
\begin{math}
\begin{tabular}{|l|r|r|}
\hline
\textbf{Systems} & \textbf{Mean Value of DAPs mutations} & \textbf{Mean Value of DPs mutations} \\
 \hline \hline
 Apache Ignite & 0.0799 & 0.0037 \\ 
 \hline
 Apache Solr & 0.1026 & 0.0232 \\
 \hline
 Eclipse & 0.1355 & 0.1128 \\
\hline
 Matsim & 0.2013 & 0.1917 \\
\hline
 Mule & 0.1989 & 0.1341 \\
\hline
 Nuxeo & 0.0848 & 0.03 \\
\hline
 oVirt & 0.0674 & 0.0252 \\
\hline
\end{tabular}
\end{math}
}
\label{tab:MeanValue}
\end{table*} 

For example, SpaghettiCode has the most representative mutation probability from Source in Mule (see Table \ref{tab:Mulemarkov}) and Blob to Sink in Apache Solr (see Table \ref{tab:SolrMarkov}). Table \ref{tab:mostrp} shows the most representative design patterns and design anti-patterns regarding mutation probabilities to/from other patterns.

\begin{table*} 
\centering
\renewcommand {\arraystretch} {1.1}
\caption{Most representative mutations between design patterns and design anti-patterns according to their mutation probabilities}
\scalebox{0.7}{
\begin{tabular}{|p{1.75cm}|l|l|l|r|}
\hline
\textbf{System} & \textbf{Mutation Type} & \textbf{From}  & \textbf{To} & \textbf{Probability} \\ \hline
\hline
\multirow{4}{*}{Apache Ignite} & DAP$\,\to\,$DAP & Blob (Bl) & AntiSingleton (AS) & 0.375\\
\cline{2-5}
& DAP$\,\to\,$DP & - & - & -\\
\cline{2-5}
  & DP$\,\to\,$DAP & - & - & -\\
  \cline{2-5}
   & DP$\,\to\,$DP & Builder (Bu) & Observer (ob) & 0.004\\
  \cline{2-5}
\hline
\multirow{4}{*}{Apache Solr} & DAP$\,\to\,$DAP & Blob (Bl) & AntiSingleton (AS) & 0.321\\
\cline{2-5}
& DAP$\,\to\,$DP & - & - & -\\
\cline{2-5}
& DP$\,\to\,$DAP & - & - & -\\
\cline{2-5}
& DP$\,\to\,$DP & FactoryMethod (FM) & Composite (Cp) & 0.012\\
\hline 
\multirow{4}{*}{Eclipe IDE} & DAP$\,\to\,$DAP & LargeClass (LC) & ComplexClass (Cc) & 0.500\\
\cline{2-5}
 & DAP$\,\to\,$DP & - & - & -\\
\cline{2-5}
 & DP$\,\to\,$DAP & FactoryMethod (FM) & LongMethod (LM) & 0.003\\
\cline{2-5}
 & DP$\,\to\,$DP & FactoryMethod (FM) & Composite (Cp) & 0.169\\
\hline 
\multirow{4}{*}{Matsim} & DAP$\,\to\,$DAP & Blob (Bl) & AntiSingleton (AS) & 0.372\\
\cline{2-5}
 & DAP$\,\to\,$DP & Blob (Bl) & FactoryMethod (FM) & 0.346\\
\cline{2-5}
 & DP$\,\to\,$DAP & Command (Cm) & SwissArmyKnife (SA) & 0.030\\
\cline{2-5}
 & DP$\,\to\,$DP & Command (Cm) & FactoryMethod (FM) & 0.387\\
\hline 
\multirow{4}{*}{Mule} & DAP$\,\to\,$DAP & Blob (bl) & AntiSingleton(AS) & 0.313\\
\cline{2-5}
 & DAP$\,\to\,$DP & RefusedParentBequest (RP) & FactoryMethod (FM) & 0.433\\
\cline{2-5}
 & DP$\,\to\,$DAP & Command (Cm) & SwissArmyKnife (SA) & 0.019\\
\cline{2-5}
 & DP$\,\to\,$DP & Command (Cm) & FactoryMethod & 0.193\\
\hline 
\multirow{4}{*}{Nuxeo} & DAP$\,\to\,$DAP & Blob (bl) & AntiSingleton(AS) & 0.283\\
\cline{2-5}
 & DAP$\,\to\,$DP & Blob (Bl) & FactoryMethod (FM) & 0.297\\
\cline{2-5}
 & DP$\,\to\,$DAP & Singleton (Si) & LazyClass (ZC) & 0.004\\
\cline{2-5}
 & DP$\,\to\,$DP & Singleton (Si) & FactoryMethod (FM) & 0.133\\
\hline 
\multirow{4}{*}{ovirt} & DAP$\,\to\,$DAP & Blob (bl) & AntiSingleton(AS) & 0.299\\
\cline{2-5}
 & DAP$\,\to\,$DP & - & - & -\\
\cline{2-5}
 & DP$\,\to\,$DAP & Singleton (Si) & AntiSingleton (AS) & 0.001\\
\cline{2-5}
 & DP$\,\to\,$DP & Singleton (Si) & Prototype (PT) & 0.097\\
\hline 
\end{tabular}
}
\label{tab:mostrp}
\end{table*} 

\paragraph{\textbf{Analysing pattern evolution}} We observe in Table \ref{tab:mostrp} that not all design patterns and design anti-pattern undergo changes. Some patterns remain stable during evolution. For example, LazyClass and MessageChain are stable design anti-patterns, while Prototype is a persistent design pattern in Apache Solr. In Matsim, design anti-patterns SwissArmyKnife, LazyClass, and MessageChain and design patterns Observer and ProtoType are stable.

However, in general, design anti-patterns tend to evolve in all studied systems. We observe that more than half of the design anti-patterns mutated into other design patterns or design anti-patterns across the different snapshots of the studied system. 

For example, in Matsim (see Table \ref{tab:MatsimMarkov}), 86\% (probability value 0.86) LongMethod remains stable and mutate with only a probability of 14\% into other patterns. In oVirt (see Table \ref{tab:Ovirtmarkov}), Blob remain persistent in the system with 39.4\% probability, while 29.9\% mutated to AntiSingleton and into other patterns with a probability of 30.8\%. As last example, in Eclipse, 45.5\% of RefusedParentBequest remain between snapshot while 27.3\% mutated to MessageChain and 27.3\% mutated into other patterns.

We saw fewer mutations among design patterns. As an example, in Apache Ignite, 97.6\% of Command remained stable, with only 0.4\% mutating into other patterns.



For a better understanding of the design pattern and design anti-pattern mutations, Figures \ref{Figure:FM-Mutations-Eclipse} to \ref{Figure:LP-Mutations-Mule} show the most representative mutations in the Markov models as graphs, for each of the systems. Because showing all possible probabilities would make the graphs unreadable, we choose a threshold of 0.100 to reduce the number of edges in each graph. The gray cells in Tables \ref{tab:EclipseMarkov} to  \ref{tab:Mulemarkov} highlight the probabilities shown in the figures. (The graphs for all the mutations in all the systems are available on-line\footnote{\url{http://www.ptidej.net/downloads/replications/emse19c/}}.)

These graphs contain information on mutating design (anti-)patterns that can help developers to avoid the mutations that could have negative impacts on software quality. 

\begin{tcolorbox}
\vspace{-0.1cm}
\textbf{Summary:} Our results show that design patterns and design anti-patterns mutate during the evolution of software systems. Despite an average of less than 20\% of the identified design anti-patterns occurrences having mutated among all the snapshots of systems, more than half of the types of design anti-patterns were involved in these mutations. For example, Table \ref{tab:MeanValue} shows that, in Matsim, 20.13\% of the design anti-pattern identified  in the first snapshot mutated during the studied period. During that period, ten types of design anti-patterns out of the 13 considered in our study were involved in mutations. In most of the systems, almost all the design patterns remained stable during evolution. Blob and Command are the design anti-patterns and design pattern with the higher mutation probabilities.
\vspace{-0.1cm}
\end{tcolorbox}

\subsection{\textbf{RQ2:} \textit{\RQTwo}}

\paragraph{\textbf{Motivation}} Knowing the causes of design (anti-)patterns mutations would be useful during maintenance. They could help developers to focus on the most frequent change types triggering patterns mutations. 


\paragraph{\textbf{Analysing change types}} We use srcML\footnote{https://www.srcml.org/} to create an XML file for each snapshot of a system and match their tags to find changed tags, as explained in Section \ref{ssec:section3.6}. We categorise change types based on our categories in Table \ref{tab:Change_types}. We compare the percentages of each change type for each system. We apply the same methodology for all the subject systems.
Figures \ref{fig:FigureChangeEclipseNew} to \ref{fig:FigureChangeMuleNew} show the types of changes per design (anti-)patterns per systems. 

Table \ref{tab:Changes} presents the numbers of each change types in all the systems. For example, in Apache Ignite, we observe that Access and  Renaming are the least and most representative change types for both design patterns and design anti-patterns. They lead to many changes in occurrences of both design anti-patterns and design patterns.

\begin{landscape}
\begin{center}
	\begin{table*}
		\renewcommand {\arraystretch} {1.05}
		\centering
		
		\caption{Number of different types of changes in design patterns and design anti-patterns\label{tab:Changes}}
		\scalebox{0.61}{
		\begin{tabular}{|p{2.7cm}|r|r|r|r|r|r|r|r|r|r|r|r|r|r|} \hline
			\textbf{Systems~$\rightarrow$} & \multicolumn{2}{c}{\textbf{Eclipse}}& \multicolumn{2}{|c}{\textbf{Nuxeo}} & \multicolumn{2}{|c}{\textbf{oVirt}}& \multicolumn{2}{|c}{\textbf{Matsim}}&\multicolumn{2}{|c}{\textbf{Apache Solr}}&\multicolumn{2}{|c}{\textbf{Apache Ignite}}&\multicolumn{2}{|c|}{\textbf{Mule}}\\ \hline 
			\multirow{2}{*}{\textbf{Change Types~$\downarrow$}}
			& 1~~~ & 2~~~ & 3~~~ & 4~~~ & 5~~~ & 6~~~ & 7~~~ & 8~~~ & 9~~~ & 10~~ & 11~~ & 12~~ & 13~~ & 14~~\\ \cline{2-15} 
			& \textbf{DAP}~~ & \textbf{DP}~~ & \textbf{DAP}~~ & \textbf{DP}~~  & \textbf{DAP}~~ & \textbf{DP}~~ & \textbf{DAP}~~ & \textbf{DP}~~ & \textbf{DAP}~~ & \textbf{DP}~~ & \textbf{DAP}~~ & \textbf{DP}~~ & \textbf{DAP}~~ & \textbf{DP}~~\\ \hline \hline
			Access & 33 & 6 & 85 & 0 & 174 & 9 & 271 & 117 & 34	& 15 & 11 & 55 & 20&12	\\ \hline
			Class & 268 & 91 & 236 & 6 & 3804 & 90 & 1697 &	690 & 721 & 155 & 781&218 & 443&198 \\ \hline
			Code block & 4197 &	1075 & 1070	& 13 & 13923 &	233 & 8805 & 3203 &	3873 & 459 & 3903&203 &1430 &413	\\ \hline 
			Comment & 32939 & 11388	& 9269	& 109 &	15013 &	411 & 22519	& 9287 & 9298 &	2616 & 14554& 1567& 6354&3780	\\ \hline 
			Control Flow & 10487 & 1870	& 966 &	10 & 6440 &	118 & 5398	& 2094 & 3166 & 636 & 3433 & 133 &916 &384	\\ \hline 
			Declaration & 9721 & 2789 &	3133 & 24 &	 27605 & 487 & 24244 & 9609 & 8803 & 1392 & 9527& 349 &3904 &1103	\\ \hline 
			Exception & 996 & 341 &	946 & 1	& 619 &	29 & 1602 &	314 & 1696 & 457 & 2076& 64& 505&173	    \\ \hline
			Import & 2566 &	835 & 2734	& 23 & 18819 & 211 & 13013 & 4679 &	4024 & 491 & 4584& 394&3234 &793	    \\ \hline
			Invocation & 1707 & 394	& 556 & 4 & 7312 & 91 &	8520 & 3026	& 2598	& 287 & 2069 & 75 &945 &240	    \\ \hline
			Method & 4060 &	942 & 1487	& 29 & 13792 & 292 & 4702 &	1922 & 3215	& 511 & 3364 & 266 & 1940&747	\\ \hline 
			Operator & 13540 & 2803	& 3533	& 8	& 35513	& 403 &	57112 &	24326 &	7963 & 702 & 7207& 525& 5241&1975	\\ \hline 
			Parameter & 5629 & 1541	& 2179	& 3 & 24488	& 292 &	22080 &	5149 & 8375 & 1069  & 9024& 332&3252 &756	\\ \hline 
			Renaming & 59254 & 14707 & 16259 & 23 &	262491 &3536 &294661&145720& 44422 & 4811 & 63961& 4396&28110&9738	\\ \hline 
			\textbf{\#Changed classes} & 10957 & 3155 & 5402& 81 & 34780 & 55 & 32596 & 13768 & 7956& 1192 & 9290& 857& 5684&2175	\\ \hline
			\textbf{Total classes} & 20331 &	7574 &	39051 &	1263 & 142537 & 2482 & 62272 & 79480 & 32332 & 5490 & 27080 & 5796 & 47146&17553	\\ \hline
			\multicolumn{15}{c}{\textbf{AP, DP}= Number of changes in design anti-patterns and design patterns respectively}\\ \hline	
		\end{tabular}
		}
	\end{table*}
	
	\begin{table*}
		\centering
		\renewcommand {\arraystretch} {1.05}
		\caption{Number of different types of changes in design patterns and design anti-patterns mutation.\label{tab:dpapMutations}}
		\scalebox{0.58}{
		\begin{tabular}{|p{2.5cm}|r|r|r|r|r|r|r|r|r|r|r|r|r|r|} \hline
			\textbf{Systems$\rightarrow$} & \multicolumn{2}{c}{\textbf{Eclipse}}& \multicolumn{2}{|c}{\textbf{Nuxeo}} & \multicolumn{2}{|c}{\textbf{oVirt}}& \multicolumn{2}{|c}{\textbf{Matsim}}&\multicolumn{2}{|c}{\textbf{Apache Solr}}&\multicolumn{2}{|c}{\textbf{Apache Ignite}}&\multicolumn{2}{|c|}{\textbf{Mule}}\\ \hline 
			\multirow{2}{*}{\textbf{Change Types~$\downarrow$}}
			& 1~~~~ & 2~~~~ & 3~~~~ & 4~~~~ & 5~~~~ & 6~~~~ & 7~~~~ & 8~~~~ & 9~~~~ & 10~~~ & 11~~~ & 12~~~ & 13~~~ & 14~~~\\ \cline{2-15} 
			 & \textbf{APDP} & \textbf{DPAP} & \textbf{APDP} & \textbf{DPAP} & \textbf{APDP} & \textbf{DPAP} & \textbf{APDP} & \textbf{DPAP} & \textbf{APDP} & \textbf{DPAP} & \textbf{APDP} & \textbf{DPAP} & \textbf{APDP} & \textbf{DPAP}\\ \hline \hline
			Access & 0 & 0 &0 & 1 &1 &1 &0 &12 &3 &0 &0 &0 &0 &1	\\ \hline
			Class & 3&1	&2	&0	&10	&0	&4 &29	&1	&7 &2 &4 &6 &3 \\ \hline
			Code block & 1 &2 &1 &4	&21	&22	&10	&132 &4	&1 &1 &3 &27 &17	\\ \hline 
			Comment & 102&192	&7	&4	&69	&31	&38	&739 &129 &242 &58 &23 &160 &85	\\ \hline 
			Control Flow & 28&38&0	&3	&7	&6	&12	&96	&44	&31	 &16 &4 &21 &3	\\ \hline 
			Declaration & 56 & 78 & 3 &2 &82 &38 &56 &412 &141 &90 &33 &16 &49 &32	\\ \hline 
			Exception & 4&6	&0	&0	&4	&1	&6	&28	&19	&13	&14 &4 &3 &2	    \\ \hline
			Import & 20&14	&3 &2	&32	&18	&28	&257 &60 &34 &17 &16 &28 &18	    \\ \hline
			Invocation & 6&	7&0	&0	&15	&20	&7	&183 &1	&6	&5 &6 & 10&3	    \\ \hline
			Method & 16&18	&2	&2	&41	&24	&7	&96	&64	&43	 &10 &12 &23 &13	\\ \hline 
			Operator & 38&37 &2	&0	&43	&66	&95	&523 &22 &20 &6 &13 &87 &32	\\ \hline 		
			Parameter & 22 & 41 & 0 & 0	&37	&20	&16	&423 &25 &47 &22 &29 &13 &22	\\ \hline 
			Renaming & 675&	469&3 &2 &297 &1036	&462 &5096	&165 &406 &100 &63 &334 &280	\\ \hline 
			\#Changed classes& 71 & 77 & 7 & 6	& 61 & 54 &58 &681 &58	&66	 &34 &29 &53 &39	\\ \hline
			\multicolumn{15}{c}{\textbf{APDP, DPAP}= Number of changes in design anti-patterns to design patterns and design patterns to design anti-patterns mutations respectively}\\ \hline
		\end{tabular}
		}
	\end{table*}
\end{center}
\end{landscape}

\begin{figure}
\centering
\scalebox{1.1}{
\begin{tikzpicture}
\pgfplotsset{every node/.append style={font=\tiny}}
\pgfplotsset{every tick label/.append style={font=\tiny}}
\begin{axis}[
    ybar,
    ymin=0,
    enlargelimits=0.2,
    legend style={at={(0.4,0.9)},anchor=north,legend columns=-1},
    bar width=1.12mm,
    ylabel={Number of Changes},
    symbolic x coords={Access, Class, Code Block, Comment, 
		Control Flow, Declaration, Exception, Import, Invocation, Method, Operator, Parameter, Renaming},
    xtick=data,
    x tick label style={rotate=45,anchor=east},
    nodes near coords,
    nodes near coords align={center},style={font=\tiny},
    every node near coord/.append style={rotate=90,anchor=south west,
    inner ysep=-1.75pt,}
    ]

\addplot [red!80!black,fill=red!40] coordinates {
(Access,33)
(Class,268) 
(Code Block,4197)
(Comment,32939)
(Control Flow,10487)
(Declaration,9721)
(Exception,996)
(Import,2566)
(Invocation,1707)
(Method,4060)
(Operator,13540)
(Parameter,5629)
(Renaming,59254)};
  
\addplot [blue!70!black,fill=blue!60] coordinates {
(Access,6) 
(Class,91)
(Code Block,1075) 
(Comment,11388)
(Control Flow,1870)
(Declaration,2789)
(Exception,341)
(Import,835)
(Invocation,394)
(Method,942)
(Operator,2803)
(Parameter,1541)
(Renaming,14707)};
\legend{Design Anti-patterns, Design Patterns}
\end{axis}
\end{tikzpicture}
}
\caption{Number of different types of changes in Eclipse classes with design anti-patterns and design patterns.}
\label{fig:FigureChangeEclipseNew}
\end{figure}

\begin{figure}
\centering
\scalebox{1.1}{
\begin{tikzpicture}
\pgfplotsset{every node/.append style={font=\tiny}}
\pgfplotsset{every tick label/.append style={font=\tiny}}
\begin{axis}[
    ybar,
    ymin=0,
    enlargelimits=0.2,
    legend style={at={(0.4,0.9)},anchor=north,legend columns=-1},
    bar width=1.12mm,
    ylabel={Number of Changes},
    symbolic x coords={Access, Class, Code Block, Comment, 
		Control Flow, Declaration, Exception, Import, Invocation, Method, Operator, Parameter, Renaming},
    xtick=data,
    x tick label style={rotate=45,anchor=east},
    nodes near coords,
    nodes near coords align={center},style={font=\tiny},
    every node near coord/.append style={rotate=90,anchor=south west,
    inner ysep=-1.75pt,}
    ]

\addplot [red!80!black,fill=red!40] coordinates {
(Access,85)
(Class,236) 
(Code Block,1070)
(Comment,9269)
(Control Flow,966)
(Declaration,3133)
(Exception,946)
(Import,2734)
(Invocation,556)
(Method,1487)
(Operator,3533)
(Parameter,2179)
(Renaming,16259)};
  
\addplot [blue!70!black,fill=blue!60] coordinates {
(Access,0) 
(Class,6)
(Code Block,13) 
(Comment,109)
(Control Flow,10)
(Declaration,24)
(Exception,1)
(Import,23)
(Invocation,4)
(Method,29)
(Operator,8)
(Parameter,3)
(Renaming,23)};
\legend{Design Anti-patterns, Design Patterns}
\end{axis}
\end{tikzpicture}
}
\caption{Number of different types of changes in Nuxeo classes with design anti-patterns and design patterns.}
\label{fig:FigureChangeNuxeoNew}
\end{figure}

\begin{figure}
\centering
\scalebox{1.1}{
\begin{tikzpicture}
\pgfplotsset{every node/.append style={font=\tiny}}
\pgfplotsset{every tick label/.append style={font=\tiny}}
\begin{axis}[
    ybar,
    ymin=0,
    enlargelimits=0.2,
    legend style={at={(0.4,0.9)},anchor=north,legend columns=-1},
    bar width=1.12mm,
    ylabel={Number of Changes},
    symbolic x coords={Access, Class, Code Block, Comment, 
		Control Flow, Declaration, Exception, Import, Invocation, Method, Operator, Parameter, Renaming},
    xtick=data,
    x tick label style={rotate=45,anchor=east},
    nodes near coords,
    nodes near coords align={center},style={font=\tiny},
    every node near coord/.append style={rotate=90,anchor=south west,
    inner ysep=-1.75pt,}
    ]

\addplot [red!80!black,fill=red!40] coordinates {
(Access,174)
(Class,3804) 
(Code Block,13923)
(Comment,15013)
(Control Flow,6440)
(Declaration,27605)
(Exception,619)
(Import,18819)
(Invocation,7312)
(Method,13792)
(Operator,35513)
(Parameter,24488)
(Renaming,262491)};
  
\addplot [blue!70!black,fill=blue!60] coordinates {
(Access,9) 
(Class,90)
(Code Block,233) 
(Comment,411)
(Control Flow,118)
(Declaration,487)
(Exception,29)
(Import,211)
(Invocation,91)
(Method,292)
(Operator,403)
(Parameter,292)
(Renaming,3536)};
\legend{Design Anti-patterns, Design Patterns}
\end{axis}
\end{tikzpicture}
}
\caption{Number of different types of changes in oVirt classes with design anti-patterns and design patterns.}
\label{fig:FigureChangeoVirtNew}
\end{figure}

\begin{figure}
\centering
\scalebox{1.1}{
\begin{tikzpicture}
\pgfplotsset{every node/.append style={font=\tiny}}
\pgfplotsset{every tick label/.append style={font=\tiny}}
\begin{axis}[
    ybar,
    ymin=0,
    enlargelimits=0.2,
    legend style={at={(0.4,0.9)},anchor=north,legend columns=-1},
    bar width=1.12mm,
    ylabel={Number of Changes},
    symbolic x coords={Access, Class, Code Block, Comment, 
		Control Flow, Declaration, Exception, Import, Invocation, Method, Operator, Parameter, Renaming},
    xtick=data,
    x tick label style={rotate=45,anchor=east},
    nodes near coords,
    nodes near coords align={center},style={font=\tiny},
    every node near coord/.append style={rotate=90,anchor=south west,
    inner ysep=-1.75pt,}
    ]

\addplot [red!80!black,fill=red!40] coordinates {
(Access,271)
(Class,1697) 
(Code Block,8805)
(Comment,22519)
(Control Flow,5398)
(Declaration,24244)
(Exception,1602)
(Import,13013)
(Invocation,8520)
(Method,4702)
(Operator,57112)
(Parameter,22080)
(Renaming,294661)};
  
\addplot [blue!70!black,fill=blue!60] coordinates {
(Access,117) 
(Class,690)
(Code Block,3203) 
(Comment,9287)
(Control Flow,2094)
(Declaration,9609)
(Exception,314)
(Import,4679)
(Invocation,3026)
(Method,1922)
(Operator,24326)
(Parameter,5149)
(Renaming,145720)};
\legend{Design Anti-patterns, Design Patterns}
\end{axis}
\end{tikzpicture}
}
\caption{Number of different types of changes in Matsim classes with design anti-patterns and design patterns.}
\label{fig:FigureChangeMatsimNew}
\end{figure}

\begin{figure}
\centering
\scalebox{1.1}{
\begin{tikzpicture}
\pgfplotsset{every node/.append style={font=\tiny}}
\pgfplotsset{every tick label/.append style={font=\tiny}}
\begin{axis}[
    ybar,
    ymin=0,
    enlargelimits=0.2,
    legend style={at={(0.4,0.9)},anchor=north,legend columns=-1},
    bar width=1.12mm,
    ylabel={Number of Changes},
    symbolic x coords={Access, Class, Code Block, Comment, 
		Control Flow, Declaration, Exception, Import, Invocation, Method, Operator, Parameter, Renaming},
    xtick=data,
    x tick label style={rotate=45,anchor=east},
    nodes near coords,
    nodes near coords align={center},style={font=\tiny},
    every node near coord/.append style={rotate=90,anchor=south west,
    inner ysep=-1.75pt,}
    ]

\addplot [red!80!black,fill=red!40] coordinates {
(Access,11)
(Class,781) 
(Code Block,3903)
(Comment,14554)
(Control Flow,3433)
(Declaration,9527)
(Exception,2076)
(Import,4584)
(Invocation,2069)
(Method,3364)
(Operator,7207)
(Parameter,9024)
(Renaming,63961)};
  
\addplot [blue!70!black,fill=blue!60] coordinates {
(Access,55) 
(Class,218)
(Code Block,203) 
(Comment,1567)
(Control Flow,133)
(Declaration,349)
(Exception,64)
(Import,394)
(Invocation,75)
(Method,266)
(Operator,525)
(Parameter,332)
(Renaming,4396)};
\legend{Design Anti-patterns, Design Patterns}
\end{axis}
\end{tikzpicture}
}
\caption{Number of different types of changes in Apache Ignite classes with design anti-patterns and design patterns.}
\label{fig:FigureChangeApacheIgniteNew}
\end{figure}

\begin{figure}
\centering
\scalebox{1.1}{
\begin{tikzpicture}
\pgfplotsset{every node/.append style={font=\tiny}}
\pgfplotsset{every tick label/.append style={font=\tiny}}
\begin{axis}[
    ybar,
    ymin=0,
    enlargelimits=0.2,
    legend style={at={(0.4,0.9)},anchor=north,legend columns=-1},
    bar width=1.12mm,
    ylabel={Number of Changes},
    symbolic x coords={Access, Class, Code Block, Comment, 
		Control Flow, Declaration, Exception, Import, Invocation, Method, Operator, Parameter, Renaming},
    xtick=data,
    x tick label style={rotate=45,anchor=east},
    nodes near coords,
    nodes near coords align={center},style={font=\tiny},
    every node near coord/.append style={rotate=90,anchor=south west,
    inner ysep=-1.75pt,}
    ]

\addplot [red!80!black,fill=red!40] coordinates {
(Access,34)
(Class,721) 
(Code Block,3873)
(Comment,9298)
(Control Flow,3166)
(Declaration,8803)
(Exception,1696)
(Import,4024)
(Invocation,2598)
(Method,3215)
(Operator,7963)
(Parameter,8375)
(Renaming,44422)};
  
\addplot [blue!70!black,fill=blue!60] coordinates {
(Access,15) 
(Class,155)
(Code Block,459) 
(Comment,2616)
(Control Flow,636)
(Declaration,1392)
(Exception,457)
(Import,491)
(Invocation,287)
(Method,511)
(Operator,702)
(Parameter,1069)
(Renaming,4811)};
\legend{Design Anti-patterns, Design Patterns}
\end{axis}
\end{tikzpicture}
}
\caption{Number of different types of changes in Apache Solr classes with design anti-patterns and design patterns.}
\label{fig:FigureChangeApacheSolrNew}
\end{figure}

\begin{figure}
\centering
\scalebox{1.1}{
\begin{tikzpicture}
\pgfplotsset{every node/.append style={font=\tiny}}
\pgfplotsset{every tick label/.append style={font=\tiny}}
\begin{axis}[
    ybar,
    ymin=0,
    enlargelimits=0.2,
    legend style={at={(0.4,0.9)},anchor=north,legend columns=-1},
    bar width=1.12mm,
    ylabel={Number of Changes},
    symbolic x coords={Access, Class, Code Block, Comment, 
		Control Flow, Declaration, Exception, Import, Invocation, Method, Operator, Parameter, Renaming},
    xtick=data,
    x tick label style={rotate=45,anchor=east},
    nodes near coords,
    nodes near coords align={center},style={font=\tiny},
    every node near coord/.append style={rotate=90,anchor=south west,
    inner ysep=-1.75pt,}
    ]

\addplot [red!80!black,fill=red!40] coordinates {
(Access,20)
(Class,443) 
(Code Block,1430)
(Comment,6354)
(Control Flow,916)
(Declaration,3904)
(Exception,505)
(Import,3234)
(Invocation,945)
(Method,1940)
(Operator,5241)
(Parameter,3252)
(Renaming,28110)};
  
\addplot [blue!70!black,fill=blue!60] coordinates {
(Access,12) 
(Class,198)
(Code Block,413) 
(Comment,3780)
(Control Flow,384)
(Declaration,1103)
(Exception,173)
(Import,793)
(Invocation,240)
(Method,747)
(Operator,1975)
(Parameter,756)
(Renaming,9738)};
\legend{Design Anti-patterns, Design Patterns}
\end{axis}
\end{tikzpicture}
}
\caption{Number of different types of changes in Mule classes with design anti-patterns and design patterns.}
\label{fig:FigureChangeMuleNew}
\end{figure}

\paragraph{\textbf{Analysing change types of mutations}} During evolution, design patterns and design anti-patterns can mutate into other design patterns and design anti-patterns. We investigate which types of changes lead to such mutations. 
Tables \ref{tab:dpapMutations} shows the number of each change types during the mutation for all the studied systems.

Results show that, in Apache Ignite, Renaming, Comment, and Declaration lead the most mutations from design anti-patterns (DAPs) to design patterns (DPs). It is almost the same for DPs-to-DAPs mutations but Parameter has more importance than Declaration. In Apache Solr and Eclipse for both DAPs-to-DPs and DPs-to-DAPs mutations, Renaming, Declaration, and Comment are the most representative change types. In Matsim, Renaming, Operator, and Declaration have the most impact on DAPs-to-DPs mutations while Renaming, Comment, and Operator lead to more DPs-to-DAPs mutations. In Mule, for both DAPs-to-DPs and DPs-to-DAPs mutations, Renaming, Comment, and Operator are the most representative change types. In Nuxeo, there are few mutations, in which Comment, Renaming, and Declaration yield DAPs-to-DPs mutations while Comment, Code Block, and Control Flow yield more DPs-to-DAPs mutations. Finally, in oVirt, Renaming, Declaration, and Comment are change types that lead to DAPs-to-DPs mutations while Renaming, Operator, and Declaration bring DPs-to-DAPs mutations.

Renaming is the most frequent change type. There are different types of renaming, described in \cite{arnaoudova2014repent}. In some types, an entity, \eg{} a package, a class, etc., is renamed. In other types, one or more terms are changed (simple and complex renaming). Sometimes, when one or more terms change, the meaning of the identifier also changes (semantic renaming). The grammar of an identifier may also change during evolution (grammar renaming). When developers use some tools to apply a renaming operation, the tool may not rename all variables consistently in all related files. Besides, certain source-code changes must be made together to preserve consistency. Changes in comments and declarations led to more mutations not as root causes of these mutations but because developers changed comments and declarations while evolving their systems for other reasons, \ie{} fixing faults. Future work include a manual, qualitative analysis of the mutations to identify their root causes.

\begin{tcolorbox}
\vspace{-0.1cm}
\textbf{Summary:} \emph{Some change types affect the mutations between design patterns and--or design anti-patterns more than others. We observe that the change types leading to mutations in all the studied systems are Renaming, Comment, Declaration, and Operator.}
\vspace{-0.1cm}
\end{tcolorbox}

\subsection{\textbf{RQ3:} \textit{\RQThree}}

\paragraph{\textbf{Motivation}} The results from RQ1 and RQ2 show that design patterns and design anti-patterns mutate during software evolution. However, they do not say anything about the impact of these mutations in software quality. Therefore, we investigate these mutations and their fault-proneness. Using this information, developers could understand the reasons of faults and take actions to reduce the risk of introducing faults.

\paragraph{\textbf{Analysing design patterns and design anti-patterns fault-proneness}} For each system, we mine its commit log and extract fault and commit IDs related to the faults and the dates when the faults were introduced. We look for faults introduced from one snapshot to the next. We use the dates to distinguish between faults appearing in one snapshot and those appearing between two extracted snapshots.

Table \ref{tab:DAP-DPfault} shows the numbers of faulty and clean of classes involved in patterns. One class can be involved in several faults in the studied snapshots so we show in Table \ref{tab:fault} the numbers of unique faulty and clean classes involved in patterns. Finally, in Table \ref{tab:faultpercentages}, we compare the percentages of faulty and clean classes involved in design patterns and design anti-patterns. Moreover, we show in Table \ref{tab:faultpercentages} the relative percentages of faults per class participating, or not, in design (anti-)patterns. 

With these tables, we summarise our whole dataset with: the numbers of (unique) faulty and non-faulty (clean) classes participating or not in design (anti-)patterns and their relative percentages. We thus can compare the prevalence of faults in different classes and confirm that classes participating in design anti-patterns have more faults than classes involved in design patterns. Thus, we have evidence supporting that classes that participate in design anti-patterns are more fault-prone than classes involved in design patterns.

For example, in Eclipse, 13.7\% of design pattern classes are fault-prone while 37.3\% of design anti-pattern classes are fault-prone. Table \ref{tab:faultpercentages} is showing similar trends in all the analysed systems.

\begin{table*} [ht]
\centering
\caption{Design anti-pattern and design-pattern mutations between faulty and clean classes}
\scalebox{0.8}{
\renewcommand{\arraystretch}{1.1}
\begin{tabular}{|l|r|r|r|}
\hline
\multirow{2}{*}{\textbf{\textbf{Systems}}} & \multicolumn{2}{c|}{\textbf{\# of Faulty classes}} & \multirow{2}{*}{\textbf{\# of Clean classes having DAPs, DPs}} \\
\cline{2-3}
 & Design Anti-patterns & Design Patterns & \\
 \hline \hline
 Apache Ignite & 10,984 & 1,051 & 81,093\\ 
 \hline
 Apache Solr & 11,156 & 219 & 109,225 \\
 \hline
 Eclipse & 15,240 & 5,182 & 19,928 \\
\hline
 Matsim & 4,053 & 1,888 & 896,510 \\
\hline
 Mule & 17,794 & 5,924 & 197,574 \\
\hline
 Nuxeo & 18,724 & 396 & 146,180 \\
\hline
 oVirt & 12,605 & 110 & 217,565 \\
\hline
\end{tabular}
}
\label{tab:DAP-DPfault}
\end{table*} 

\begin{table*} [ht]
\centering
\caption{Faulty and clean classes}
\scalebox{0.7}{
\renewcommand{\arraystretch}{1.1}
\begin{tabular}{|l|rr|rr|rr|rr|rr|}
\hline
\multirow{2}{*}{\textbf{Systems}} & \multicolumn{4}{c|}{\textbf{\# of Faulty Classes}} & \multicolumn{2}{c|}{\textbf{\# of Faulty Classes}} & \multicolumn{4}{c|}{\textbf{\# of Clean Classes}} \\
\cline{2-11}
& \multicolumn{2}{c|}{DAPs} & \multicolumn{2}{c|}{DPs} & \multicolumn{2}{c|}{\textbf{without Patterns}} & \multicolumn{2}{c|}{DAPs} &
\multicolumn{2}{c|}{DPs} \\
\hline \hline
Apache Ignite & 10,984 & (685) & 1,051 & (84) & 3,984 & (3,984) & 71,784 & (3,474) & 9,309 & (395)\\
\hline
Apache Solr & 11,156 & (638) & 219 & (22) & 6,351 & (6,351) & 101,044 & (3,447) & 8,181 & (392)\\
\hline
Eclipse & 15,240 & (591) & 5,182 & (217) & 12,285 & (12,285) & 13,610 & (562) & 6,318 & (213)\\
\hline
Matsim & 4,053 & (469) & 1,888 & (115) & 291 & (291) & 326,460 & (14,425) & 570,050 & (9,913)\\
\hline
Mule & 17,794 & (2,955) & 5,924 & (196) & 4,370 & (4,370) & 126,980 & (7,341) & 70,594 & (1,593)\\
\hline
Nuxeo & 18,724 & (1,487) & 396 & (68) & 7,414 & (7,414) & 143,276 & (3,659) & 2,904 & (170)\\
\hline
oVirt & 12,605 & (377) & 110 & (8) & 523 & (523) & 214,027 & (7,653) & 3,538 & (214)\\
\hline
\end{tabular}
}
\label{tab:fault}
\end{table*} 

\begin{table*} [ht]
\centering
\caption{Faulty and clean classes in percentages}
\scalebox{0.7}{
\renewcommand{\arraystretch}{1.1}
\begin{tabular}{|l|r|r|r|r|}
\hline
\multirow{2}{*}{\textbf{Systems}} & \multicolumn{2}{c|}{\textbf{Design anti-patterns}} & \multicolumn{2}{c|}{\textbf{Design patterns}} \\
\cline{2-5}
& Faulty Classes (\%) & Clean Classes (\%) &
Faulty Classes (\%) & Clean Classes (\%) \\
\hline \hline
Apache Ignite & 14.7\% & 74.9\% & 1.8\% & 8.5\% \\
\hline
Apache Solr & 14.1\% & 76.6\% & 0.48\% & 8.7\% \\
\hline
Eclipse & 37.3\% & 35.5\% & 13.7\% & 13.4\% \\
\hline
Matsim & 1.9\% & 57.9\% & 0.46\% & 39.7\% \\
\hline
Mule & 24.4\% & 60.7\% & 1.6\% & 13.2\% \\
\hline
Nuxeo & 27.6\% & 67.9\% & 1.3\% & 3.2\%\\
\hline
oVirt & 4.6\% & 92.7\% & 0.09\% & 2.6\%\\
\hline
\end{tabular}
}
\label{tab:faultpercentages}
\end{table*}

\paragraph{\textbf{Analyzing mutations fault-proneness}} A mutation between design patterns and design anti-patterns can lead to faults. We use clean and faulty classes and their participation (or not) into design patterns and design anti-patterns to identify the mutations experienced by these faulty classes.

Table \ref{tab:TransFault} presents the most representative mutations that led to faults in each studied system. We observe that mutations from design anti-patterns to other design anti-patterns are more faulty. LongParameterList to LongMethod or LongMethod to LazyClass are such mutations in Apache Ignite. 

In Eclipse, Matsim, and Mule, there are mutations from design patterns to design patterns that also led to more faults. FactoryMethod to Decorator in Eclipse, Builder to FactoryMethod in Matsim and Mule are such mutations. 

There are also mutations from design anti-patterns to design patterns that led to faults as well, like AntiSingleton to FactoryMethod in Matsim or FactoryMethod to LongMethod in Eclipse. 


\begin{table*} [ht]
\centering
\caption{Most representative mutations between design patterns and design anti-patterns according to their mutation probabilities and fault-proneness}

\scalebox{0.8}{
\renewcommand {\arraystretch} {1.1}
\begin{tabular}{|l|l|l|l|r|}
\hline
\textbf{System} & \textbf{Mutation Type} & \textbf{From}  & \textbf{To} & \textbf{Probability} \\ \hline
\hline
\multirow{2}{*}{Apache Ignite} & 
 DAP$\,\to\,$DAP & LongParameterList & LongMethod & 0.571\\
\cline{2-5}
& DAP$\,\to\,$DAP & LongMethod & LazyClass & 0.285\\
\cline{2-5}
\hline
\multirow{3}{*}{Apache Solr} &  DAP$\,\to\,$DAP & RefusedParentBequest & MessageChain & 0.427\\
\cline{2-5}
& DAP$\,\to\,$DAP & LongMethod & LazyClass & 0.156 \\
\cline{2-5}
& DAP$\,\to\,$DAP & ComplexClass & ClassDataShouldBePrivate & 0.156\\
\cline{2-5}
\hline 
\multirow{3}{*}{Eclipe IDE} &  
 DP$\,\to\,$DP & FactoryMethod & Decorator & 0.492\\
 \cline{2-5} 
  & DAP$\,\to\,$DAP & LongMethod & LazyClass & 0.385\\
\cline{2-5}
 & DP$\,\to\,$DAP & FactoryMethod & LongMethod & 0.056\\
\cline{2-5}
\hline 
\multirow{3}{*}{Matsim} & 
DP$\,\to\,$DP & Builder & FactoryMethod & 0.677\\
\cline{2-5}
& DAP$\,\to\,$DAP & SpagettiCode & RefusedParentBequest & 0.152\\
\cline{2-5}
 & DAP$\,\to\,$DP & AntiSingleton & FactoryMethod & 0.114\\
\cline{2-5}
\hline 
\multirow{3}{*}{Mule} & 
DP$\,\to\,$DP & Builder & FactoryMethod & 0.479 \\
\cline{2-5}
& DAP$\,\to\,$DP & ComplexClass & FactoryMethod & 0.264\\
\cline{2-5}
& DAP$\,\to\,$DAP & ComplexClass & ClassDataShouldBePrivate & 0.223\\ 
\cline{2-5}
\hline  
\multirow{2}{*}{Nuxeo} & DAP$\,\to\,$DAP & LazyClass & LargeClass & 0.285\\
\cline{2-5}
 & DP$\,\to\,$DP & Singleton & FactoryMethod & 0.495\\
\hline 
\multirow{2}{*}{oVirt} & DAP$\,\to\,$DAP & Blob & AntiSingleton & 0.722\\ 
\cline{2-5}
 & DP$\,\to\,$DP & Singleton & Prototype& 0.166\\ 
\hline 
\end{tabular}
}
\label{tab:TransFault}
\end{table*} 

\begin{tcolorbox}
\vspace{-0.1cm}
\textbf{Summary:} \emph{We observed that in some systems, as expected and shown in previous work, design anti-patterns are more fault-prone than design patterns. We also showed that some mutations are more fault-prone than others, in particular mutations from design anti-patterns to design patterns or to other design anti-patterns.}
\vspace{-0.1cm}
\end{tcolorbox}

\subsection{\textbf{RQ4:} \textit{\RQFour}}

\paragraph{\textbf{Motivation}} Different types of changes have different impacts on the software systems due to their differences in functionality and the ripple effects of changes. Some types of changes likely introduce more faults than others. Thus, understanding which types of changes increase the fault-proneness of the mutations could help developers to foresee and prevent faults by preventing/planning such changes during software evolution.

\paragraph{\textbf{Analysing change types leading to faults}} We use the same data as in RQ2 and RQ3. For each system, we identify the number of faulty classes that have changed through mutations between design anti-patterns and--or design patterns. Table \ref{tab:ChangeAn2De} shows the number of change types that led to faults. We report that, in all studied systems, Renaming, Comment, and Operator are the change types that lead to more faults.

\begin{table*}[ht]
\centering
\caption{Numbers of change types in the studied systems leading to faults} \label{tab:noFaultyChanges}
\scalebox{0.71}{
\renewcommand{\arraystretch}{1.1}
\begin{tabular}{|p{2.1cm}|r|r|r|r|r|r|r|}\hline
\textbf{Systems $\rightarrow$ }& \textbf{Apache Ignite} & \textbf{Apache Solr} & \textbf{Eclipse~~~} & \textbf{Matsim~~~} & \textbf{Mule~~~} & \textbf{Nuxeo~~~} & \textbf{oVirt~~~}\\\hline
\textbf{\textbf{Change Types}} &\# changes  &\# changes&\# changes &\# changes & \# changes &\# changes &\# changes\\
\hline \hline
  Access &18  &18  & 37 & 22 & 11 & 62 & 25\\ \hline
  Class &689  &431  &306  &306  & 422 &208  &763 \\ \hline
  Code block & 2,972 & 2,102 & 4,919 & 1,284 &1,306  & 854 & 3,833\\ \hline
  Comment & 12,169 & 5,498 & 38,150 & 2,813 & 6,270 &6,161  &4,653 \\ \hline
  Control flow &2,903  & 1,935 & 11,678 & 660 &  1067& 819 &2,406 \\ \hline
  Declaration &5,912  &  5,386& 11651 &  3,129& 3,191 & 2,628 &7,484 \\ \hline
  Exception &1,696  & 1221 & 1255 & 140 &  526& 786&  210\\ \hline
  Import & 2,831 &2,400  &2,958  & 1,425 &2,443  & 2,064 & 4,268 \\ \hline
  Invocation &1,550  &1,196  & 1,986 & 1,061 & 840 & 476 &1,882 \\ \hline
  Method & 2,509 & 1,851 & 4,093 &637  & 1,697 & 1,229 & 3,619\\ \hline
  Operator &5,120  & 4,364 & 16,094 & 6,215 & 4,228 & 2,675 & 8,134\\ \hline
  Parameter & 6,418 & 3,504 & 5,108 & 2,655 & 2,607 & 1,815 & 5,337 \\ \hline
  Renaming & 47,811 & 24,640 & 67,040 & 32,445 & 22,968 &  12,736& 65,245 \\ \hline
  \textbf{Total changed classes} &  5,505& 4,163 & 11,934 & 3,073 & 4,324 & 3,514 &7,150 \\ \hline
\end{tabular}
}
\label{tab:ChangeAn2De}
\end{table*} 

\paragraph{\textbf{Analyzing fault-proneness of classes with design patterns and design anti-patterns}} Table \ref{tab:Changefault} presents the numbers of faulty changed classes and Figures \ref{fig:DPFaultyChangedClasses} and \ref{fig:APFaultyChangedClasses} show the percentages of faulty changed classes participating in design patterns and design anti-patterns for all the systems. Figure \ref{fig:AP_DPFaultyChangedClasses} compares the numbers of faulty and clean classes changed in all the snapshots of the studied systems. Each first bar presents the number of faulty and clean changed classes participating in design anti-patterns and the second one is faulty and clean classes having design patterns. We observe that change types have impacts on the fault-proneness of changed classes. Changed classes participating in design patterns are less faulty than those participating only in design anti-patterns. 

Figures \ref{fig:DPFaultyChangedClasses} and \ref{fig:APFaultyChangedClasses} show that some of the faulty classes are those which had changed in the past. For example, in Eclipse, the percentages of faulty classes participating in design patterns is 81\% while for those participating in design anti-patterns it is 86\%. The differences between these two categories are more visible in Apache Solr, where 51\% of changed classes are participating in design anti-patterns, and only 11\% of them have design patterns. In Rhino, changes impact fault-proneness significantly, because, on average, more than 85\% of changed classes are faulty. Thus, the trend is that changed classes with design anti-patterns tend to be more fault-prone than changed classes with design patterns.

\begin{table*} [ht]
\centering
\caption{Numbers of faulty and clean changed classes}
\scalebox{0.9}{
\renewcommand{\arraystretch}{1.1}
\begin{tabular}{|l|l|r|r|}
\hline
\textbf{Systems} & \textbf{Patterns}  & \textbf{\# Faulty classes} & \textbf{\# Clean classes}\\
\hline \hline
\multirow{2}{*}{Apache Ignite} & Design Anti-patterns & 5,112 & 4,178\\ 
 \cline{2-4}
 & Design Patterns & 393 & 464\\
\hline
\multirow{2}{*}{Apache Solr} & Design Anti-patterns & 4,035 & 3,921\\
\cline{2-4}
 & Design patterns & 128 & 1,064\\
\hline
\multirow{2}{*}{Eclipse} & Design Anti-patterns & 9,406 & 1,551\\
\cline{2-4}
 & Design patterns & 2,554 & 601\\
\hline
\multirow{2}{*}{Matsim} & Design Anti-patterns & 2,549 & 30,042\\
\cline{2-4}
 & Design patterns & 524 & 13,244\\
\hline
\multirow{2}{*}{Mule} & Design Anti-patterns & 3,374 & 2,311\\
\cline{2-4}
 & Design patterns & 950 & 1,225\\
\hline
\multirow{2}{*}{Nuxeo} & Design Anti-patterns & 3,469 & 1,935\\
\cline{2-4}
 & Design patterns & 45 & 36\\
\hline
\multirow{2}{*}{oVirt} & Design Anti-patterns & 7,075 & 27,705\\
\cline{2-4}
 & Design patterns & 75 & 482\\
\hline
\end{tabular}
}
\label{tab:Changefault}
\end{table*}

\begin{figure}[ht]
\centering
\scalebox{0.8}{
  \begin{tikzpicture}
  \begin{axis}[
      every node near coord/.style={
     },
     legend style={
       font=\footnotesize,
       cells={anchor=east},
       legend columns=5,
       at={(0.50,-0.30)},
       anchor=north,
       /tikz/every even column/.append style={column sep=0.1cm}
     },
    title={},
    ybar stacked, ymin=0,
    bar width=3mm,
    ylabel={Percentages of changed classes},
    symbolic x coords=     {Apache Ignite-DP, Apache Solr-DP, Eclipse-DP, Matsim-DP, Mule-DP, Nuxeo-DP, ovirt-DP},
    xtick=data,
    x tick label style={rotate=45,anchor=east},
]
  \addplot [fill=red] coordinates {
  ({Apache Ignite-DP},0.51)
  ({Apache Solr-DP},0.11)
  ({Eclipse-DP},0.81)
  ({Matsim-DP},0.04)
  ({Mule-DP},0.43)
  ({Nuxeo-DP},0.55)
  ({ovirt-DP},0.13)
};
    \addplot [fill=yellow] coordinates {
  ({Apache Ignite-DP},0.49)
  ({Apache Solr-DP},0.89)
  ({Eclipse-DP},0.19)
  ({Matsim-DP},0.96)
  ({Mule-DP},0.57)
  ({Nuxeo-DP},0.45)
  ({ovirt-DP},0.87)
};
  \legend {Design patterns faulty classes, Clean classes} 
  \end{axis}
  \end{tikzpicture}
  }
 \caption{Faulty changed classes percentages with design pattern in the studied systems} 
\label{fig:DPFaultyChangedClasses}
\end{figure}
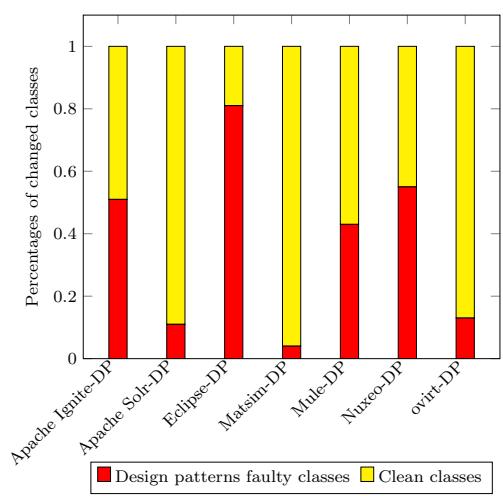

\begin{figure}[ht]
\centering
\scalebox{0.8}{
  \begin{tikzpicture}
  \begin{axis}[
      every node near coord/.style={
     },
     legend style={
       font=\footnotesize,
       cells={anchor=east},
       legend columns=5,
       at={(0.50,-0.30)},
       anchor=north,
       /tikz/every even column/.append style={column sep=0.1cm}
     },
    title={},
    ybar stacked, ymin=0,
    bar width=3mm,
    ylabel={Percentages of changed classes},
    symbolic x coords=     {Apache Ignite-DAP, Apache Solr-DAP, Eclipse-DAP, Matsim-DAP, Mule-DAP, Nuxeo-DAP, ovirt-DAP},
    xtick=data,
    x tick label style={rotate=45,anchor=east},
]
  \addplot [fill=blue] coordinates {
  ({Apache Ignite-DAP},0.55)
  ({Apache Solr-DAP},0.51)
  ({Eclipse-DAP},0.86)
  ({Matsim-DAP},0.08)
  ({Mule-DAP},0.59)
  ({Nuxeo-DAP},0.64)
  ({ovirt-DAP},0.20)
};

  \addplot [fill=yellow] coordinates {
  ({Apache Ignite-DAP},0.45)
  ({Apache Solr-DAP},0.49)
  ({Eclipse-DAP},0.14)
  ({Matsim-DAP},0.92)
  ({Mule-DAP},0.41)
  ({Nuxeo-DAP},0.36)
  ({ovirt-DAP},0.80)
};
  \legend {Design anti-patterns faulty classes, Clean classes} 
  \end{axis}
  \end{tikzpicture}
  }
 \caption{Faulty changed classes with design anti-patterns percentages in the studied systems} 
\label{fig:APFaultyChangedClasses}
\end{figure}
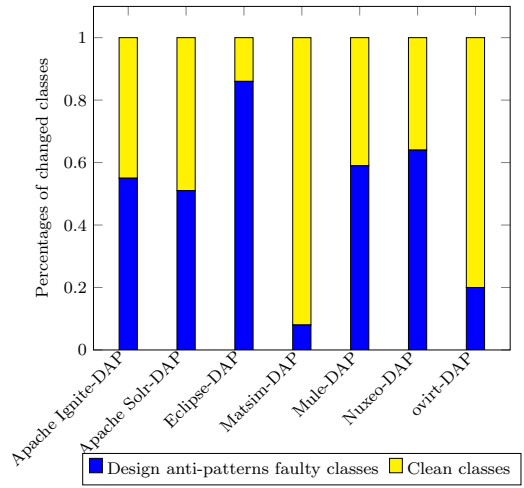

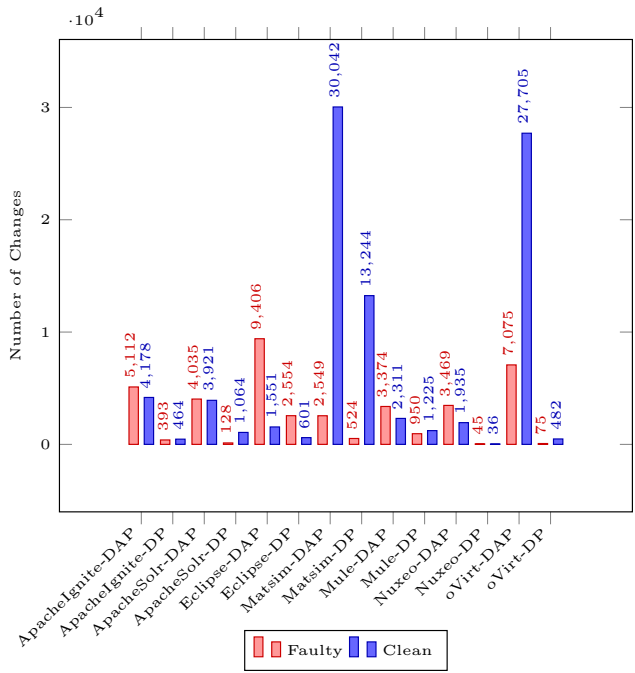
\begin{figure}
\centering
\scalebox{1.1}{
\begin{tikzpicture}
\pgfplotsset{every node/.append style={font=\tiny}}
\pgfplotsset{every tick label/.append style={font=\tiny}}
\begin{axis}[
    ybar,
    ymin=0,
    enlargelimits=0.2,
    legend style={at={(0.5,-0.25)},anchor=north,legend columns=-1},
    bar width=1.12mm,
    ylabel={Number of Changes},
    symbolic x coords={ApacheIgnite-DAP, ApacheIgnite-DP, ApacheSolr-DAP, ApacheSolr-DP, Eclipse-DAP, Eclipse-DP, Matsim-DAP, Matsim-DP, Mule-DAP, Mule-DP, Nuxeo-DAP, Nuxeo-DP, oVirt-DAP, oVirt-DP},
    xtick=data,
    x tick label style={rotate=45,anchor=east},
    nodes near coords,
    nodes near coords align={center},style={font=\tiny},
    every node near coord/.append style={rotate=90,anchor=south west,
    inner ysep=-1.75pt,}
    ]

\addplot [red!80!black,fill=red!40] coordinates {
(ApacheIgnite-DAP,5112) (ApacheIgnite-DP,393) (ApacheSolr-DAP,4035) (ApacheSolr-DP,128) (Eclipse-DAP,9406) (Eclipse-DP,2554) (Matsim-DAP,2549) (Matsim-DP,524) (Mule-DAP,3374) (Mule-DP,950) (Nuxeo-DAP,3469) (Nuxeo-DP,45) (oVirt-DAP,7075) (oVirt-DP,75)};
  
\addplot [blue!70!black,fill=blue!60] coordinates {
(ApacheIgnite-DAP,4178) (ApacheIgnite-DP,464) (ApacheSolr-DAP,3921) (ApacheSolr-DP,1064) (Eclipse-DAP,1551) (Eclipse-DP,601) (Matsim-DAP,30042) (Matsim-DP,13244) (Mule-DAP, 2311) (Mule-DP,1225) (Nuxeo-DAP,1935) (Nuxeo-DP,36) (oVirt-DAP,27705) (oVirt-DP,482)};
\legend{Faulty, Clean}
\end{axis}
\end{tikzpicture}
}
\caption{Faulty changed classes with design anti-patterns and design patterns in the studied systems} 
\label{fig:AP_DPFaultyChangedClasses}
\end{figure}

\begin{tcolorbox}
\vspace{-0.1cm}
\textbf{Summary:} \emph{Some change types applied to design patterns and design anti-patterns make software systems more fault-prone compared to others. We observed that, in all the studied systems, Renaming, Comment, and Operator are the change types from design patterns to design anti-patterns that most lead to faults.}
\vspace{-0.1cm}
\end{tcolorbox}
\section{Discussion}
\label{sec:Discussion}

By comparing the results between all studied systems, we observed remarkable differences in the proportions and types of mutations of design anti-patterns and design patterns. For example, in oVirt in Table \ref{tab:Ovirtmarkov}, 29.9\% of Blob mutated to AntiSingleton (the highest mutation percentages). However, we did not observe in the same system any mutation from Blob to a design pattern. On the contrary, in Nuxeo, although the mutation of Blob to AntiSingleton remains frequent (28.3\%), the highest mutation proportion observed is Blob to Factory Method (29.7\%).

Another remarkable observation is that, in all studied systems, the occurrences of the design anti-pattern SwissArmyKnife and of the design pattern Prototype never mutated. We cannot draw any conclusion about a potential impact of the mutation of these patterns on the quality of the systems. 


Different software systems may have different design patterns and--or design anti-patterns and may evolve differently. From our analysis, we observe different mutation behavior for all analyzed systems because these systems have different designs, contexts, and development teams. We observed that some design patterns and--or design anti-patterns remained unchanged in all releases and they did not mutate during evolution. 

For example, class \emph{org.mule.test.infrastructure.process.MuleUtils} in all the snapshots of Mule, is a LongMethod design anti-pattern. This design anti-pattern is introduced when developers continue adding new functionalities to a method while nothing is never taken out. Usually, developers prefer to add code to an existing method instead of creating a new one \cite{brown1998antipatterns}, which means that another line is added and then another, giving birth to a tangle of spaghetti code. This longer method or function become harder to understand and maintain.

We found that some of the design anti-patterns are mutated frequently to design patterns when developers are correcting faults during the evolution of the systems. Blob is the most mutated design anti-pattern in Apache Ignite. It mutated to AntiSingleton with 37.5\% probability. Blob presents a single class with a large number of attributes, operations, etc., surrounded by a number of data classes. A Blob is too complex for reuse and testing, while such classes are inefficient, and expensive to load into memory. 

There are also some design patterns that mutated to design anti-patterns. Command is an example of a design pattern that often mutated into SwissArmyKnife (38.7\% of the time) in Matsim. SwissArmyKnife is an excessively complex class interface. Developers attempted to provide for all possible uses of the class. They added a large number of interfaces (APIs) to meet all possible needs. The code to create separate Command classes grow to encompass more functionalities and become a SwissArmyKnife.

We observed that the change types that led to more mutations are Renaming, Comment, Operator, and Declaration, in most of the studied systems. These types of changes helped developers to correct their systems and remove some design anti-patterns. We also noticed that the most frequent mutation was LongParameterList to LongMethod.

We found in some analysed systems, that design anti-patterns are more fault-prone than classes involved in design patterns, while in some other systems, it is quite the opposite. Although these observations mean that employing design patterns may not always benefit the software quality, and introducing anti-patterns will not systemically compromise the software quality, some previous research related similar observations.

Bieman \etal{} \cite{bieman2003design} observed that large classes participating to design patterns were more change-prone. Voka\v{c} \etal{} \cite{vokavc2004defect} found a significant differences in the fault-proneness of different design patterns. Gatrell \etal{} \cite{gatrell2009design} observed that pattern-based classes are more change-prone and less stable than non-pattern classes \cite{vokavc2004defect}. Long \cite{long2001software} showed the benefits of some anti-patterns in the context of code reuse.

The context of using design anti-patterns or design patterns \cite{long2001software}, their static and historical relationships \cite{jaafar2013analysing,jaafar2013mining}, and their internal characteristics, such as their numbers of lines of code  \cite{bieman2003design}, could be compounding factors that lead to our previous observations, which confirm previous findings that software quality decreases with design anti-patterns and increases with design patterns.

A fault is an error that causes a software system to produce an incorrect or unexpected result or to behave in unintended ways. Although several previous works, \eg{} \cite{khomh2012exploratory,jaafar2014anti}, showed that an important proportion of faults may be related to design patterns, anti-patterns, and mutations among them, many other faults remain unrelated to patterns or their mutations. Table \ref{tab:fault} shows the numbers of (unique) faults per classes participating or not in some design (anti-)patterns. It shows that, in all studied systems except Matsim, the numbers of classes with faults but not participating in any design (anti-)pattern is higher than the number of classes with faults and participating in some design (anti-)pattern.

The results in this paper generally confirm results from previous studies and provide new research directions to further understand the impact of programming practices, like patterns and evolution, on software quality. First, while we confirmed that, generally, classes participating in design anti-patterns have more faults than classes participating in design patterns or no patterns; we also reported contradictory cases, like the design anti-pattern SwissArmyKnife and the design pattern Prototype. Thus, further studies are needed to understand the root causes of such cases, through manual, qualitative analyses of the changes leading to the faults.

We showed that some mutations are more frequent than others between some design anti-patterns and design patterns. We reported changes that led, concretely, to these mutations. However, our study is only on co-occurrences of these changes and mutations: more studies are required to understand why some patterns mutate into others in some systems but not others. We believe that these mutations and their differences may be due to the systems themselves and their developers but further studies are required to identify root causes and assert causation.

Finally, we showed that some mutations are more fault prone than others, which could guide developers but also researchers. For developers, our observations can help avoiding harmful changes, \ie{} fault-prone mutations, and focus some refactoring activities on beneficial changes. For researchers, our observations provide a first step in understanding the introduction of patterns and the changes that may lead to them as well as their impact on fault-proneness.
\section{Threats to Validity}
\label{sec:Treats to Validity}

We now discuss potential threats to the validity of the results of our study, following existing guidelines \cite{yin2013case,wohlin2012experimentation}.

\paragraph{Construct validity} These threats concern the relation between theory and observation. We know that the used design pattern and design anti-pattern detection techniques (DECOR and DeMIMA) in this study include some subjective understanding related to the definition of design patterns and design anti-patterns. Their authors reported recall rates of 100$\%$ for both techniques while the precision in the worst case was 31$\%$. We accept that the precision of these techniques is a concern. Some false positive classes may pass the validation because they ``looks like'' playing a role in some patterns.

We also accept that, in finding change types which led to faults, we could have matched classes that are not representing the actual same class. For example, class \texttt{C} is not match with \texttt{a.b.C.java} but could be matched with the \texttt{b.c.java}. Moreover, we know that during evolution, class names change as well. As for precision, the manual validation could be affected by subjectiveness or human error.  We should consider each type of renaming as we may misinterpret that there is a mutation between design patterns and design anti-patterns, while in fact the class name changed and the patterns remained stable.

\paragraph{Internal validity} This threat concerns factors affecting our results. This threat is about the causality drawn from the study. It concerns our selection of studied systems and methodology. The accuracy of DECOR and DeMIMA impacts our results, because the number of design patterns and design anti-patterns computed with DECOR and DeMIMA is used to calculate the probabilities of mutations. Other detection techniques should be used to validate our findings.

Our results show correlations between design anti-patterns and design patterns, their mutations, and faults. However, they do not show causation. Hence, it is possible that some of the changes, which led to mutations, \eg{} changes to comments, although correlated to mutations, are not the root causes of these mutations. Identifying these root causes would require studying each change leading to mutations individually, manually, which is future work.

\paragraph{Conclusion validity} These threats concern the relationship between the treatment and the results. We paid attention in choosing the systems.

We used the SZZ algorithm \cite{sliwerski2005changes} to identify commits introducing faults. Although this algorithm may yield false positive results, it has been successfully employed in previous works, such as \cite{kamei2013large,fukushima2014empirical}. In this paper, to increase the algorithm's accuracy, we removed all fault-inducing commit candidates that only changed blank or comment lines. Moreover, the static analysis tool, srcML, can identify about 100 types of code elements from source code.

To make our results more actionable for software practitioners, we manually grouped similar element tags into 12 major change types as shown in Table~\ref{tab:Change_types}, which can help developers carefully change and review fault-prone code.

\paragraph{Reliability Validity} These threats concern the possibility of replicating the study. We provide all the necessary data on-line\footnote{\url{http://www.ptidej.net/downloads/replications/emse19c/}} to help other researchers replicate our work.

\paragraph{External validity} These threats concern the ability to generalize our results. We studied seven software systems with different sizes, domains, and complexity. We selected only Java systems because of the tools. We also chose some of these systems because they have been used in previous studies. Their numbers of lines of code range from hundred of thousands to several millions. These systems are widely used and have active developers community. They have several years of evolution histories. They are available on-line. However, all of them are written in Java and are open source. In the future, we plan to investigate more diverse set of systems. Moreover, we also want to study larger projects, with other programming languages, such as C++. 

We analysed commits instead of releases to cover as much as possible the whole histories of these systems. We choose thirteen design anti-patterns and eight design patterns among the many available patterns.
\section{Conclusion and Future Work}
\label{sec:Conclusion and Future Work}

We investigated the evolution and impacts of design patterns and design anti-patterns in terms of change- and fault-proneness during software evolution. We built Markov models to analyse the mutations of design patterns and design anti-patterns in seven open-source Java systems: Apache Ignite, Apache Solr, Eclipse, Matsim, Mule, Nuxeo, and Ovirt. We identified the change types that led to mutations and we calculated the probabilities of all possible mutations. Finally, we reported which patterns are mostly mutated into which other patterns (including appearance and disappearance) as well as change types.

Results showed that design patterns and design anti-patterns mutate into one another during software evolution and that these mutations impact the fault-proneness of classes participating in these patterns. Generally, when a mutation led to the introduction of a design anti-pattern, quality in terms of fault-proneness decreased; when a design anti-pattern was removed or mutated into a design pattern,  quality increased.

Using this information, developers can focus on the design patterns that are most likely to mutate into design anti-patterns and--or to have more faults. Thus, this information can help evolution and quality assurance by focusing refactoring efforts on classes with design (anti-)patterns that could mutate into patterns with higher fault-proneness.

In future work, we intend to apply our study on more systems written in different programming languages and also consider more design (anti-)patterns. We will also attempt to identify the root causes of mutations by studying manually and qualitatively the changes leading to mutations.

\balance
\newcommand{\student}[1]{#1}
\newcommand{\uml}{UML}
\bibliographystyle{IEEEtran} \scriptsize
\def\IEEEbibitemsep{0pt}
\bibliography{BibiTex}

\end{document}